\documentclass[a4paper,11pt]{article}
\pdfoutput=1 % if your are submitting a pdflatex (i.e. if you have
\usepackage{jheppub} % for details on the use of the package, please
\usepackage{amsmath, amsfonts}

\usepackage[T1]{fontenc} % if needed
\usepackage{xcolor}
\usepackage{subfigure}
\usepackage{stmaryrd}
\usepackage{enumerate}
\usepackage{tikz}
\usetikzlibrary{calc,intersections,positioning,arrows}

\newcommand{\beq}{\begin{equation}}
\newcommand{\eeq}{\end{equation}}
\newcommand{\bea}{\begin{eqnarray}}
\newcommand{\eea}{\end{eqnarray}}

\newcommand{\ep}{\epsilon}

\newcommand{\nn}{\nonumber}

\newcommand{\MKK}{M_{\mathrm{KK}}}

\DeclareMathOperator{\Tr}{Tr}

\usepackage{color}

\usepackage[normalem]{ulem}  % \sout{old text} for strikeout

\newcommand{\non}{\nonumber\\}

\newcommand{\be}{\begin{equation}}
\newcommand{\ee}{\end{equation}}

\title{Thermal pion condensation: holography meets lattice QCD}
\author[a]{Nicolas Kovensky}
\author[b]{and Andreas Schmitt}

\affiliation[a]{Instituto de F\'isica de La Plata - CONICET
Diagonal 113 e/ 63 y 64, 1900 - La Plata, Argentina.}
\affiliation[b]{Mathematical Sciences and STAG Research Centre, University of Southampton, Highfield Campus, Southampton
SO17 1BJ, United Kingdom.}
\emailAdd{nicolas.kovensky@iflp.unlp.edu.ar}
\emailAdd{a.schmitt@soton.ac.uk}

\abstract{The holographic Witten-Sakai-Sugimoto model  is often employed to describe strongly-coupled baryonic and isospin-asymmetric matter, for example in the context of neutron stars. Here we consider the case of vanishing baryon chemical potential, where detailed comparisons to data from lattice QCD are possible. To this end, we extend previous works by including a realistic pion mass and pion condensation into the decompactified limit of the model and evaluate the system for arbitrary isospin chemical potentials and temperatures. After suitably fixing the 3 parameters of the model, we find that the overall phase structure is in excellent agreement with lattice results. This also holds for observables at low temperatures in the strongly coupled regime, while we discover and discuss some discrepancies at large temperatures. Our findings give  reassurance for the validity of previous and future applications of  this model and highlight the aspects where improvements are needed.}

\begin{document} 
\maketitle
\flushbottom

\section{Introduction}
\label{sec:intro}

Understanding the phases of Quantum Chromodynamics (QCD) is crucial for describing a wide variety of physical phenomena, ranging from quark-gluon plasma dynamics in heavy-ion collisions to the cosmological evolution of the early universe, and the physics of neutron stars, including their mergers. Ideally, we would like to have control over the thermodynamics of QCD in a large -- and multi-dimensional -- parameter space. Of particular interest are baryon and isospin chemical potentials and the temperature (other possible variables are the magnetic field and a strangeness chemical potential). As a computational tool for describing strong interactions, QCD is, however, notoriously difficult to handle away from the perturbative regime at ultra-high temperatures and/or densities. The so-called sign problem further renders the possibility of simulating baryonic matter on the lattice extremely challenging. Hence, one must often resort either to an appropriate effective field theory description, for example chiral perturbation theory ($\chi$PT) \cite{GasserLeutwyler:1983CPT,Ecker:1994gg}, or to phenomenological models, such as the linear sigma model \cite{Gell-Mann:1960mvl,Lee:1968da} or the Nambu-Jona--Lasinio  model  \cite{Nambu:1961tp,Nambu:1961fr}, or to the gauge-gravity duality (``holography'') \cite{Maldacena:1997re,Gubser:1998bc,Witten:1998qj}. Holography provides results for the strong-coupling limit of a gauge theory via classical gravity in a higher-dimensional curved space. Since the gravity dual of QCD is not known, this gauge theory is different from, but ideally as close as possible to, QCD.  

Here we employ the holographic Witten-Sakai-Sugimoto (WSS) model \cite{Witten:1998zw,Sakai:2004cn,Sakai:2005yt}, which, in a certain -- albeit inaccessible -- limit is dual to QCD at a large number of colors $N_c$. We focus on two-flavor matter, $N_f=2$, at nonzero temperature and isospin chemical potential, but at vanishing baryon chemical potential. In this scenario \cite{Son:2000xc,Son:2000by,Splittorff:2000mm},  comparisons to lattice QCD \cite{Kogut:2002tm,Kogut:2002zg,Brandt:2017oyy,Brandt:2018omg,Cuteri:2021hiq,Brandt:2022hwy,Abbott:2023coj,Abbott:2024vhj} are possible. Although without direct applications to the above mentioned physical systems, isospin QCD has therefore become an ideal playground for testing different theoretical approaches to dense QCD matter, most importantly in the intermediate density regime, which lies outside the realm of applicability of both (leading-order) $\chi$PT and perturbative QCD \cite{Kojo:2024sca}. Additionally, there are interesting open questions in isospin QCD itself. For instance, the precise way in which the mesonic degrees of freedom are progressively replaced by those associated to deconfined quarks as the density increases is yet to be fully understood. And, moreover, as exploited for instance in refs.\ \cite{Hidaka:2011jj,Fujimoto:2023unl,Abbott:2024vhj}, isospin QCD can provide bounds on QCD with baryon chemical potential via general inequalities for the QCD partition function \cite{Cohen:2003ut}. 

The WSS model is a string-theoretic (``top-down'') holographic construction, in which $N_c$ D4-branes characterize the glue sector of the non-supersymmetric, non-conformal, and thus QCD-like dual field theory. At large $N_c$, they backreact on the bulk ambient space, thus generating a non-trivial gravitational background. Flavor physics are described in terms of $N_f$ pairs of D8- and $\overline{\rm D8}$-branes, accounting for left- and right-handed quarks and thus providing a geometric realization of chiral symmetry and the spontaneous breaking thereof. Assuming $N_f \ll N_c$, the flavor branes can be studied in the probe limit, which is employed in the vast majority of applications, including the present  work. In the context of dense and/or hot matter the model has been frequently used and improved over the years, see for instance refs.\
\cite{Bergman:2007wp,Parnachev:2007bc,Aharony:2007uu,Rebhan:2008ur,Preis:2010cq,Preis:2011sp,Preis:2012fh,Bigazzi:2014qsa,Li:2015uea,BitaghsirFadafan:2018uzs,Kovensky:2021kzl,Bartolini:2022gdf,Bartolini:2023wis}, and we build upon various of these improvements, in particular those of refs.\
\cite{Kovensky:2019bih,Kovensky:2020xif,Kovensky:2021ddl,Kovensky:2023mye}. 

Meson condensation generated by an isospin chemical potential was first investigated in this model in ref.\ \cite{Aharony:2007uu}. Our study  improves on this work in several ways. Firstly, we work with a nonzero pion mass $m_\pi$. Compared to other holographic constructions, introducing a pion mass is somewhat more complicated in the WSS model, and we do so with the help of an effective mass term as in refs.\
\cite{Kovensky:2019bih,Kovensky:2020xif,Kovensky:2023mye}, following the proposal of refs.\
\cite{Aharony:2008an,Hashimoto:2008sr,Argyres:2008sw}. An alternative approach based on including a tachyonic mode was proposed and employed in refs.~\cite{Casero:2007ae,Bergman:2007pm,Dhar:2007bz,Dhar:2008um}. In ref.~\cite{McNees:2008km} it was argued that both effective descriptions account for the same open-string physics. Therefore, although a rigorous connection between the two approaches has not been made, we expect them to lead to similar results, at least in the regime of small (realistic) pion masses. Secondly, we consider a non-antipodal configuration of the flavor branes. This renders their embedding dynamical, as it takes into account the backreaction of the medium. This effect allows us to go beyond 
$\chi$PT, which emerges in the low-energy limit of the model in the {\it antipodal} setting, where the flavor branes are fixed and follow geodesics \cite{Sakai:2004cn}. We shall indeed see that close to the vacuum our results approach those of $\chi$PT, while in general they capture effects beyond $\chi$PT. Thirdly, we assume the asymptotic separation of the flavor branes $L$ to be small compared to the inverse of the so-called Kaluza-Klein mass $M_{\rm KK}$. In this ``decompactified'' limit, where we only work in the deconfined background geometry, the thermodynamics depend nontrivially on temperature. The price we have to pay is that we cannot describe the gluonic deconfinement phase transition. However, this version of the model leads to a much richer phase structure, which otherwise would require us to go beyond the probe brane approximation, which is very difficult\footnote{One solution to this problem is to work at large $N_c$ \textit{and} $N_f$, with $N_f/N_c$ fixed, which is done in the (somewhat more phenomenological) V-QCD model \cite{Jarvinen:2011qe,Alho:2012mh}. }.  Setting $N_c=3$ in our eventual results, there are only 3 relevant model parameters to be fixed, namely $m_\pi$, $L$, and the 't Hooft coupling of the boundary theory $\lambda$\footnote{In this version of the model, the value of $M_{\rm KK}$  is irrelevant for all observables we consider.}, in  contrast to comparable field-theoretic phenomenological models, which typically have a much larger number of parameters. 

Our approach in particular generalizes our own previous work, where pion condensation was considered in the deconfined geometry, but in the chiral limit \cite{Kovensky:2021ddl} and in the presence of a pion mass, but in the confined geometry with antipodal separation of the flavor branes \cite{Kovensky:2023mye}. For the main purpose of this paper -- the comparison with lattice QCD -- working with a realistic pion mass {\it and} in the decompactified limit is crucial. Our results also improve other holographic studies of pion condensation at nonzero isospin chemical potential, performed in various ``bottom-up'' approaches  \cite{Albrecht:2010eg,Nishihara:2014nva,Cao:2020ske}.

Our paper is organized as follows. The model is introduced in section \ref{sec:2}. The general setup, in particular the mass term and the identification of the chiral and pion condensates, are explained in  section \ref{sec:setup}, while the relevant equations for the pion-condensed phase are derived in section \ref{sec:pi-phase}.
(Phases where the pion condensate vanishes are straightforward generalizations of previous works and are thus relegated to appendix \ref{AppA}.) In section \ref{sec:fit}, we explain our parameter fit, and the results, which require numerical evaluation, are presented and discussed in section \ref{sec:results}. We focus on zero temperature in section \ref{sec:results-zeroT}, before we analyze temperature effects in section \ref{sec:results-Teffects}. The phase diagram in the plane of temperature and isospin chemical potential is shown in section \ref{sec:results-phasediagram}, including the identification of the regime where the speed of sound is larger than its value in the conformal limit. We summarize our findings and give an outlook in section  \ref{sec:outlook}.

%%%%%%%%%%%%%%%%%%%%%%%%%%%%%%%%%%%%%%%%%%%%%%%%%%%%%%%%%
\section{Thermal pion condensation in holographic QCD}
\label{sec:2}
%%%%%%%%%%%%%%%%%%%%%%%%%%%%%%%%%%%%%%%%%%%%%%%%%%%%%%%%%
Here we introduce the holographic model employed in this paper, in a setup that allows us to work at nonzero temperature $T$ and isospin chemical potential $\mu_I$, while the baryon chemical potential is turned off throughout. 
In particular, we derive the holographic description of thermal pion condensation at nonzero quark masses, and identify a medium-dependent pion decay constant.

%%%%%%%%%%%%%%%%%%%%%%%%%%%%%%%%%%%%%%%%%%%%%%%%%%%%%%%%%
\subsection{Setup}
\label{sec:setup}
%%%%%%%%%%%%%%%%%%%%%%%%%%%%%%%%%%%%%%%%%%%%%%%%%%%%%%%%%

\begin{figure} [t]
\begin{center}
\includegraphics[width=\textwidth]{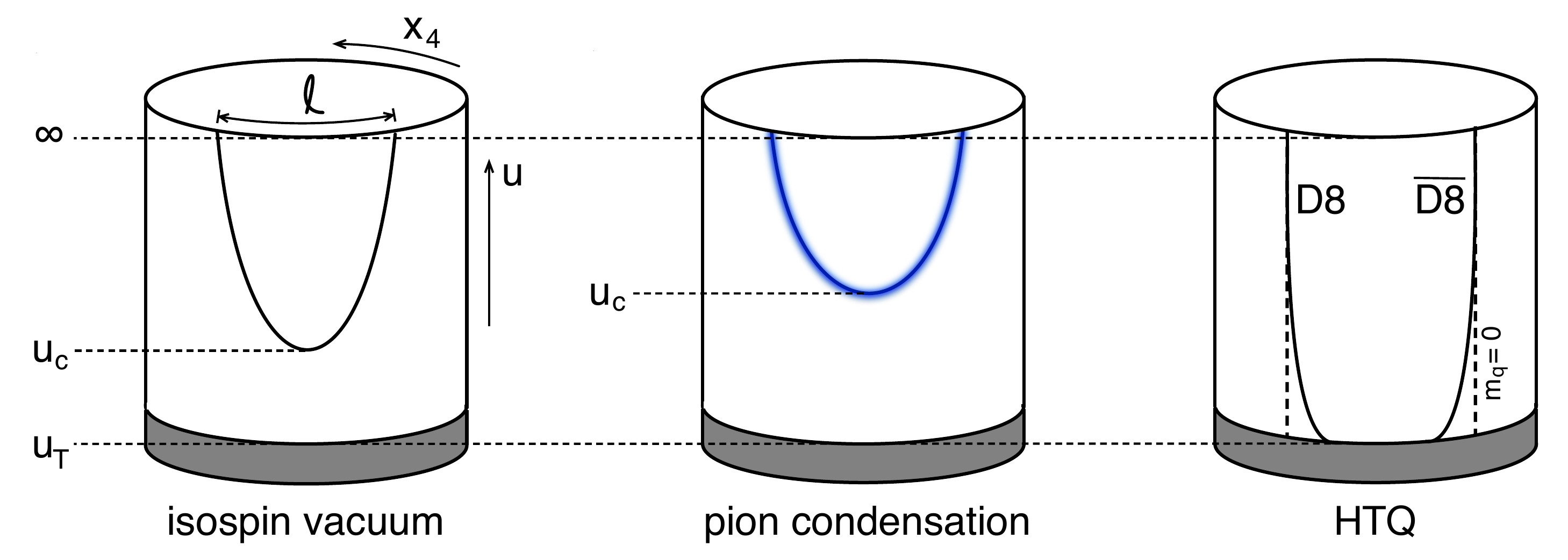}
\caption{Bulk geometry of the most relevant phases. In the dimensionless units used in our calculation, the flavor branes are asymptotically separated by a distance $\ell=LM_{\rm KK}$ in the compact $x_4=X_4 M_{\rm KK}$ direction. In the bulk, their embedding in the fixed background geometry is dynamically computed, including the point $u_c$ where they connect if chiral symmetry is broken. The radial (``holographic'') coordinate is denoted by $u\in[u_T,\infty]$, where $u_T$ depends on temperature and denotes the location of the black hole horizon. Pion condensation is distinguished from the isospin vacuum by a different gauge field configuration on the flavor branes, here simply indicated by a colored embedding. We work with a nonzero quark mass $m_q$, in which case the branes are non-straight even if they stretch down all the way to the horizon in the high-temperature quark phase (HTQ). The low-temperature quark phase (LTQ), which contains string sources,  turns out to be energetically disfavored in the phase diagram we consider and is thus not shown here.}  
\label{fig:embeddings}
\end{center}
\end{figure}

The WSS model is based on type-IIA superstring theory compactified on a circle, whose coordinate is usually denoted by $X_4$, with radius $\MKK^{-1}$ and periodicity conditions for the fermions that break supersymmetry completely. The gravitational background is sourced by $N_c$ D4-branes and thus accounts for the gluonic sector of the dual field theory, with a Hawking-Page phase transition at the Kaluza-Klein scale associated with deconfinement. Flavor is added by means of $N_f$ pairs of D8-$\overline{\rm D8}$-branes, which, in the probe brane approximation, do not affect the background geometry (either confined or deconfined). The global chiral symmetry group of the boundary field theory corresponds to a U($N_f$)$\times$ U($N_f$) gauge symmetry on the flavor branes. The branes  are separated by a distance $L$ at the holographic boundary. Whether they connect in the bulk is a dynamical question. If they do, chiral symmetry is spontaneously broken. In the original version of the model, where the branes are maximally separated, $L=\pi\MKK^{-1}$, chiral symmetry breaking is, within the probe brane approximation, locked to the deconfinement transition for all chemical potentials. Here we consider $L \ll \pi \MKK^{-1}$, which unlocks deconfinement and chiral symmetry restoration. 
In this limit, it depends on temperature {\it and} the chemical potentials whether the flavor branes remain separated in the bulk or not. Additionally, pion condensation in the chirally broken phase is accounted for by certain boundary conditions of the gauge fields on the flavor branes, as we shall explain in detail. Therefore, we will dynamically account for a chirally broken phase without pion condensation, the pion-condensed phase, and a chirally restored phase; see figure \ref{fig:embeddings} for an illustration of the corresponding geometries.  

We shall work with the following action for the embedding function of and gauge fields on the flavor branes, 
\begin{equation}
   S =  S_{\rm DBI} + S_{q}+S_m \, .
\end{equation}
Here, $S_{\rm DBI}$ is the Dirac-Born-Infeld (DBI) action, $S_q$ contains string sources \cite{Bergman:2007wp}, and $S_m$ is the effective mass term \cite{Aharony:2008an,Hashimoto:2008sr,McNees:2008km,Kovensky:2019bih}. We will now define and explain each term separately, using the dimensionless coordinates and variables introduced in ref.\ \cite{Li:2015uea}.

%%%%%%%%%%%%%%%%%%%%%%%%%%%%%%%%%%%%%%%%%%%%%%%%%%%%%%%%%
\subsubsection{DBI action}
\label{sec:DBI}
%%%%%%%%%%%%%%%%%%%%%%%%%%%%%%%%%%%%%%%%%%%%%%%%%%%%%%%%%

For our purposes, we may use the DBI action in the deconfined geometry from ref.\ \cite{Kovensky:2021ddl}, where we drop the abelian gauge field and the non-abelian spatial components,  
\bea
\label{SDBI}
S_{\rm DBI} = {\cal N}N_f\frac{V}{T}\int_{u_c}^\infty du\, u^{5/2}\zeta(u) \, , 
\eea
with
\bea\label{zetadef}
\zeta(u) \equiv \sqrt{1+u^3f_T(u)x_4'(u)^2-K_a'(u)K_a'(u)}  \, , 
\eea
where $K_a(u)$, $a=1,2,3$, are the (dimensionless) temporal components of the non-abelian SU($N_f$) gauge fields (following the  notation of ref.\ \cite{Kovensky:2023mye}), $x_4(u)$ describes the probe brane embedding, the prime denotes the derivative with respect to $u$, $V$ is the three-volume, and $f_T(u)$ is the blackening factor of the background metric, 
\be
    f_T(u) = 1 - \frac{u_T^3}{u^3} \, ,
\ee
where the location of the horizon $u_T$ is related to the temperature and its dimensionless counterpart $t$ as 
\be\label{tdef}
t \equiv  \frac{T}{\MKK} =  \frac{3}{4\pi}\sqrt{u_T} \, .  
\ee
The (dimensionful) prefactor in the DBI action is 
\begin{equation} \label{normN}
    \mathcal{N} = \frac{N_c \MKK^4 \lambda_0^3}{6\pi^2} \, ,
\end{equation}
where 
\be
\lambda_0 \equiv \frac{\lambda}{4\pi} \, .
\ee
The isospin chemical potential will be encoded in the boundary conditions for the fields $K_a(u)$. We will thus work with a dimensionless isospin chemical potential $\bar{\mu}_I$ whose relation to its physical counterpart is determined by the definition of the dimensionless gauge fields \cite{Li:2015uea},
\be \label{muIdim}
\mu_I = \lambda_0 M_{\rm KK}\bar{\mu}_I \, .
\ee
In eq.~\eqref{SDBI} we have assumed the branes to be connected at $u=u_c$ (see left and middle geometries in figure \ref{fig:embeddings}). The same form of the DBI action with $u_c$ replaced by $u_T$ is used if the branes reach the horizon (see right geometry in figure \ref{fig:embeddings}). More details on the DBI action can be found for instance in ref.\ \cite{Kovensky:2021ddl}, where additional gauge fields are included to account for baryon number. On the other hand, this reference only included a single non-abelian temporal component $K_3(u)$, which is sufficient in the chiral limit. In the presence of a pion mass, as first discussed in ref.\ \cite{Kovensky:2023mye}, all three components $K_a(u)$ are necessary to implement pion condensation, at least in the approach that makes use of a convenient chiral rotation that is also explained in detail in ref.\ \cite{Kovensky:2023mye} and which will become clearer below when we discuss the pion-condensed phase explicitly.   

%%%%%%%%%%%%%%%%%%%%%%%%%%%%%%%%%%%%%%%%%%%%%%%%%%%%%%%%%
\subsubsection{String sources}
\label{sec:string}
%%%%%%%%%%%%%%%%%%%%%%%%%%%%%%%%%%%%%%%%%%%%%%%%%%%%%%%%%

It is conceivable to include string sources in the case of connected flavor branes, stretching from $u_T$ to the tip of the branes at $u_c$, thus creating a cusp in the embedding. They give rise to a contribution of deconfined quarks and anti-quarks to the isospin density. This is a direct generalization of the string sources introduced in the context of baryon number. In that case, the resulting configurations are unstable in the chiral limit \cite{Bergman:2007wp}, but do play a crucial role in the low-temperature regime of the phase diagram for nonzero pion mass \cite{Kovensky:2019bih}. The contribution to the action is (see also ref.\ \cite{Kovensky:2020xif})
\bea
S_q = {\cal N}N_f\frac{V}{T}\int_{u_c}^\infty du\,\bar{n}_I [u-u_T-K_3(u)]\delta(u-u_c) \, .
\eea
We shall only consider string sources in the absence of pion condensation, such that the only contribution to the isospin density comes from the strings. Therefore, here it {\it is} sufficient to only include a single component $K_3(u)$, with the isospin chemical potential as a boundary condition. We have introduced the dimensionless isospin density $\bar{n}_I$, which is related to its dimensionful counterpart by 
\be \label{nIq}
n_{I} = \frac{N_cN_f\lambda_0^2M_{\rm KK}^3}{6\pi^2}\bar{n}_{I} \, .
\ee
(This relation holds for all phases we consider, not only for the isospin density from string sources.) We shall find that the phase with string sources -- the isospin analogue of the low-temperature quark phase (LTQ) of ref.\ \cite{Kovensky:2019bih} -- is not preferred anywhere in the phase diagram. Consequently, for our main results, $S_q$ will play no direct role.

%%%%%%%%%%%%%%%%%%%%%%%%%%%%%%%%%%%%%%%%%%%%%%%%%%%%%%%%%
\subsubsection{Mass contribution and condensates}
\label{sec:mass}
%%%%%%%%%%%%%%%%%%%%%%%%%%%%%%%%%%%%%%%%%%%%%%%%%%%%%%%%%

We write the effective mass term in the action as  
\begin{equation}
\label{Smdef}
    S_m = \int_0^{1/T} d\tau\int d^3 x \, {\cal L}_m = - \mathcal{N} N_f \frac{V}{T} \frac{A \cos\theta}{2\lambda_0} \, , 
\end{equation}
where $A$ is only nonzero in the presence of a pion mass, and where  
$\theta$ encodes the pion condensate. The notation follows earlier works where a pion mass and/or a pion condensate was included
\cite{Kovensky:2019bih,Kovensky:2020xif,Kovensky:2023mye}. However, this work is the first that deals with the pion condensate away from the chiral limit {\it and} a nontrivial embedding of the flavor branes that responds to pion condensation. Therefore, let us explain in detail how the Lagrangian ${\cal L}_m$ comes about and how we can extract medium-dependent chiral and pion condensates from it.  

First we recall that massive quarks are difficult to implement in the WSS model mainly because there is no spatial direction available for the usual geometric realization of the Higgs mechanism by separating the flavor branes. Quark masses can nevertheless be included in an effective way by an open Wilson line  stretching between D8- and  $\overline{{\rm D8}}$-branes \cite{Aharony:2008an,Hashimoto:2008sr,McNees:2008km}. 
In the strong coupling limit, the corresponding expectation value $\langle\mathcal{O}\rangle$ must be proportional to the exponential of the on-shell action $S_{\rm WS}$ of the resulting worldsheet instanton between the flavor branes, which receives contributions from the Nambu-Goto action $S_{\rm NG}$ and from a boundary term $S_{\partial}$, 
\begin{equation}
    \langle\mathcal{O}\rangle = c e^{-S_{\rm WS}}= c e^{-S_{\rm NG} + S_{\partial}} \, , 
\end{equation}
where $c$ is a constant, to be fixed below. 
The boundary term gives the chiral field 
\begin{equation} \label{Uchiral}
    U = e^{-S_\partial} =  \exp \left(i \int_{-\infty}^{+\infty} dz \, a_z\right) \, ,   
\end{equation}
with the (dimensionless) gauge field $a_z$. Here, the radial coordinate $z$ parameterizes both halves of the connected flavor branes and is defined via $u^3 = u_c^3+ u_c z^2$. In the DBI action (\ref{SDBI}), we have already set $a_z=0$ as a particular gauge choice. This gives a trivial chiral field, $U=1$, and it appears we cannot describe pion condensation. However, there is nothing wrong with a general, nontrivial chiral field in this subsection since the mass term $S_m$ is invariant under global chiral rotations (a transformation of $U$ is undone by the transformation of the mass matrix). Working later with $a_z=0$ in the practical calculation is possible if the chiral rotation that renders $U$ trivial is taken into account in the boundary conditions of the gauge fields $K_a(u)$ \cite{Aharony:2007uu,Rebhan:2008ur,Kovensky:2021ddl,Kovensky:2023mye}. With this in mind we may temporarily keep $a_z$ nonzero and $U$ nontrivial,
\bea \label{angles}
U &=&\frac{\sigma+i\pi_a\tau_a}{f_\pi}=\cos\psi\cos\theta+i\cos\psi\sin\theta(\tau_1\cos\varphi+\tau_2\sin\varphi)+i\tau_3\sin\psi \, , 
\eea
where $\tau_a$ are the Pauli matrices, $f_\pi$ is the pion decay constant, and $\sigma^2\equiv f_\pi^2-\pi_a\pi_a$. We will work with the parameterization $(\psi,\theta,\varphi)$, but have included in Eq.\ (\ref{angles}) an alternative parameterisation to make the connection to the usual pion fields $\pi_a$ explicit: $\psi$ and $\theta$ are nonzero for neutral and charged pion condensation, respectively, and $\varphi$ can be understood as the phase of the complex field associated with the charged pions. Our system is symmetric with respect to $\varphi$ and thus this angle will drop out of all physical results. 

The Nambu-Goto action is \cite{Kovensky:2019bih}
\be
\label{SNG}
S_{\rm NG} = -2\lambda_0\left[\phi_T(u_c)x_4(u_c)+\int_{u_c}^\infty du\, \phi_T(u) x_4'(u)\right]\, , 
%\qquad S_{\rm NG}^0 = -\frac{\lambda_0}{\ell}\pi\tan\frac{\pi}{16} \, ,
\ee
where 
\begin{equation}\label{phiT}
    \phi_T(u) \equiv \int \frac{du}{\sqrt{f_T(u)}} = \frac{u}{\sqrt{f_T(u)}} \left\{
    1-\frac{3u_T^3}{4 u^3 f_T^{1/6}(u)} {}_2F_1 \left[
    \frac{1}{6},\frac{2}{3},\frac{5}{3}, -\frac{u_T^3}{u^3 f_T(u)}
    \right]\right\} \, , 
\end{equation}
with the hypergeometric function ${}_2F_1$. 
This action can be used for  connected branes [where $x_4(u_c)=0$] and for branes that reach the horizon [where $u_c$ needs to be replaced by $u_T$ and $x_4(u_T)$ is a dynamical quantity]. The expression in eq.~\eqref{SNG} is already renormalized by subtracting the vacuum contribution \cite{Kovensky:2019bih}.   

The Lagrangian of the effective mass term is constructed by multiplying the expectation value for the Wilson line with the quark mass $m_q=m_u\simeq m_d$, 
\be \label{O}
{\cal L}_m = - \frac{m_q}{2}\Tr[\langle{\cal O}\rangle + \langle{\cal O}^\dag\rangle] 
%= -\frac{m_q}{2}ce^{-S_{\rm NG}} \Tr[U+U^\dag]
=-2m_qce^{-S_{\rm NG}} \cos\theta  \, , 
\ee
where we have used the parameterization (\ref{angles}) and set the neutral pion field to zero, $\psi=0$. This  Lagrangian has the same form as the lowest-order mass term in $\chi$PT. However, importantly, the Nambu-Goto action depends on temperature explicitly and is sensitive to the shape of the flavor branes via the function $x_4(u)$, which in turn depends implicitly on temperature and chemical potential as well as on the quark mass. In contrast, $c$ must not depend on the medium. 

The (medium-dependent) chiral condensate $\langle\bar{\psi}\psi\rangle$ is defined through 
the derivative of an effective potential (here given by the on-shell action) with respect to the corresponding source, the quark mass. Analogously, we can also introduce a source for the pion condensate $\langle\pi^\pm\rangle$ \cite{Adhikari:2020kdn}. These sources appear explicitly only in ${\cal L}_m$, and we can ignore implicit dependencies since the total action will be minimized with respect to all quantities that depend implicitly on the sources. Therefore, we obtain 
\begin{subequations} \label{psipsipi}
\bea \label{qqc}
\langle\bar{\psi}\psi\rangle &=& -c e^{-S_{\rm NG}}\cos\theta \, , \\[2ex] 
\langle\pi^\pm \rangle &=& -c e^{-S_{\rm NG}}\sin\theta \, .
\eea
\end{subequations}
We can now fix the proportionality constant $c$. To this end, we evaluate eq.\ (\ref{qqc}) in the vacuum $T=\mu_I=0$ where there is no pion condensation, $\theta=0$,
\be \label{qqc0}
\langle\bar{\psi}\psi\rangle_0 = -c e^{-S_{\rm NG}^0} \, .
\ee
We insert this result into our mass term and expand in powers of the pion field from Eq.\ (\ref{angles}) to read off the vacuum mass of the pion from the quadratic term. This relates the chiral condensate to the pion mass,
\begin{equation}\label{GOR}
- m_q\langle\bar{\psi}\psi\rangle_0 = \frac{m_\pi^2 f_\pi^2}{2} \, ,
\end{equation}
which is the well-known Gell--Mann-Oakes-Renner relation \cite{Gell-Mann:1968hlm}. We thus obtain
\be
c= \frac{f_\pi^2m_\pi^2}{2m_q}\frac{1}{e^{-S_{\rm NG}^0}} \, . 
\ee
 Assuming a homogeneous system, the integration in eq.\ (\ref{Smdef}) over imaginary time $\tau$ and position space simply gives a factor $V/T$, and we can read off 
\be \label{A} 
A = \frac{2\alpha}{\lambda_0^2}e^{-S_{\rm NG}} \, , 
\ee
where, following the notation of refs.\
\cite{Kovensky:2019bih,Kovensky:2020xif}, we have introduced 
\be \label{alphadef}
\alpha \equiv \frac{3\pi^2f_\pi^2m_\pi^2}{N_cM_{\rm KK}^4e^{-S_{\rm NG}^{0}}}  \, . 
\ee
The quantities $A$ and $\alpha$ are a convenient way to deal with the mass effects self-consistently in the practical calculation. While $\alpha$ will be considered as a model parameter ($\alpha=0$ in the chiral limit), $A$ will be used as a dynamical variable we have to solve for by coupling the equations of motion with eq.\ (\ref{A}).  

In the deconfined geometry of the WSS model, the pion decay constant is 
\be\label{Fpim}
 F_\pi^2 \equiv 
 \frac{{\cal N}}{2\lambda_0^2M_{\rm KK}^2} \left[\int_{u_c}^\infty du\,  \frac{\zeta(u)}{u^{5/2}}\right]^{-1} \, ,
\ee
with $\zeta(u)$ from eq.\ (\ref{zetadef}).  
In this expression, we allow the integral to depend on the medium and the pion mass. If $\zeta(u)$ is evaluated in the  
vacuum and at zero pion mass we recover the known result  \cite{Kovensky:2020xif,Callebaut:2011ab}
\be 
F_\pi^2(T=\mu_I=m_\pi=0) = \frac{128 \lambda_0 N_c M_{\rm KK}^2}{3\pi\ell^3} \left(\frac{\Gamma[9/16]}{\Gamma[1/16]}\right)^3\frac{\Gamma[11/16]}{\Gamma[3/16]}\, , \label{fpi0}
\ee
where
\be
\ell \equiv LM_{\rm KK} \, .
\ee
Importantly, the pion decay constant introduced in eq.\ (\ref{GOR}) does depend on the pion mass,
\be \label{fpiFpi}
f_\pi \equiv F_\pi(T=\mu_I=0) \, .
\ee
We emphasize that this is true for $\langle\bar{\psi}\psi\rangle_0$ and $S_{\rm NG}^0$ as well, i.e., these are vacuum quantities evaluated at nonzero pion mass\footnote{This is different from ref.\ \cite{Kovensky:2019bih}, where $S_{\rm NG}^0$ denotes the Nambu-Goto action in the vacuum and at {\it zero} pion mass.}. This is necessary to reproduce the correct zero-temperature onset of pion condensation, as we shall see in section \ref{sec:pi-phase}.

%%%%%%%%%%%%%%%%%%%%%%%%%%%%%%%%%%%%%%%%%%%%%%%%%%%%%%%%%
\subsubsection{Possible phases}
\label{sec:phases}
%%%%%%%%%%%%%%%%%%%%%%%%%%%%%%%%%%%%%%%%%%%%%%%%%%%%%%%%%

The setup explained in the previous subsections allows us to consider the following phases: 

\begin{itemize}

\item {\it Isospin vacuum.} Here, pion condensate and isospin density are zero, and chiral symmetry is spontaneously broken. This phase is identical to the phase termed {\it mesonic} in ref.\ \cite{Kovensky:2019bih}, which in turn is a generalization to nonzero pion mass of the chirally broken phase in the deconfined geometry already discussed in early applications of the WSS model \cite{Aharony:2006da,Horigome:2006xu}. 

\item {\it Low-temperature quark phase (LTQ).} In this phase the pion condensate is zero, and isospin density is generated solely by string sources. At nonzero pion mass, the LTQ phase replaces (for low temperatures) the chirally restored phase with straight and disconnected branes, which would be favored in the chiral limit. The LTQ phase was discussed in ref.\ \cite{Kovensky:2019bih} in the context of a baryon chemical potential and the generalization to isospin chemical potential is straightforward.

\item {\it High-temperature quark phase (HTQ).} Here, the pion condensate also vanishes and isospin number is created by quarks and anti-quarks because the flavor branes reach the horizon, i.e., without explicit string sources. This phase was also first discussed for nonzero baryon chemical potential \cite{Kovensky:2019bih} and reduces to the chirally restored phase with straight branes if the pion mass goes to zero. It only exists for sufficiently large temperatures.  

\item {\it Pion-condensed phase}. This phase, where the pion condensate and thus also the isospin density are nonzero, is constructed for the first time in this paper, building on simpler versions in the confined geometry and/or in the chiral limit \cite{Aharony:2007uu,Rebhan:2008ur,Kovensky:2021ddl,Kovensky:2023mye}. 
We set the string sources to zero in this phase.

\end{itemize}

The treatment of the first three phases is straightforward. The isospin vacuum can be directly taken from ref.\ \cite{Kovensky:2019bih}, while the LTQ and HTQ phases are easily derived from their analogues at nonzero baryon chemical potential. For the sake of a self-contained presentation we briefly go through the main equations of all three phases, but defer them to appendix \ref{AppA}. The pion-condensed phase, however, will be discussed in detail in the next subsection.

One may ask whether it makes any sense to allow for string sources to coexist with pion condensation, which would constitute a fifth phase in our list. Such a phase would presumably contain isospin number from quarks as well as from the pion condensate, not unlike the holographic quarkyonic phase, where baryon number is generated by quarks from string sources as well as from baryons \cite{Kovensky:2020xif}. In fact, it is tempting to speculate whether such a phase allows for a holographic realization of the so-called BEC-BCS crossover, where a phase of ``molecules'', the pions, forming a Bose-Einstein condensate (BEC), gradually transitions to a phase of weakly coupled Cooper pairs according to the Bardeen-Cooper-Schrieffer (BCS) theory \cite{progress}.

%%%%%%%%%%%%%%%%%%%%%%%%%%%%%%%%%%%%%%%%%%%%
\subsection{Pion condensation}
\label{sec:pi-phase}
%%%%%%%%%%%%%%%%%%%%%%%%%%%%%%%%%%%%%%%%%%%%

%%%%%%%%%%%%%%%%%%%%%%%%%%%%%%%%%%%%%%%%%%%%
\subsubsection{Boundary conditions} 
%%%%%%%%%%%%%%%%%%%%%%%%%%%%%%%%%%%%%%%%%%%%

The dynamical fields in the pion-condensed phase are $x_4(u)$ and $K_a(u)$. Let us first discuss their boundary conditions. Since the flavor branes join at $u_c$, we may set $x_4(u_c)=0$ and $x_4(\infty) = \ell/2$, or
\be
\label{ell}
\frac{\ell}{2} = \int_{u_c}^\infty du\, x_4' \,. 
\ee
The isospin chemical potential might be straightforwardly implemented in the ultraviolet boundary condition of $K_3$. However, as already mentioned below eq.\ (\ref{Uchiral}), pion condensation is most conveniently discussed in the gauge $a_z=0$, which requires a chiral rotation that affects all components of the non-abelian gauge potentials. A suitable rotation for the case of a nonzero pion mass was discussed in ref.\ \cite{Kovensky:2023mye} (the rotation applied in the chiral limit \cite{Aharony:2007uu,Rebhan:2008ur,Kovensky:2021ddl} would lead to asymmetric gauge fields with respect to $z\to -z$ if the pion mass is nonzero, which is less convenient). Since the arguments of ref.\ \cite{Kovensky:2023mye} regarding the chiral rotation are obviously independent of the background geometry, we may impose the boundary conditions 
\begin{subequations} \label{K123inf}
\bea
K_1(\infty) &=& -\bar{\mu}_I\sin\theta\sin\varphi \,  , \\[2ex]
K_2(\infty) &=& \bar{\mu}_I\sin\theta\cos\varphi \,  , \\[2ex]
K_3(\infty) &=& \bar{\mu}_I\cos\theta \,  ,
\eea
\end{subequations}
and 
\begin{equation}
K_1(u_c)=K_2(u_c)=K_3'(u_c)=0 \, .    
\end{equation}
The infrared boundary value $K_3(u_c)$ needs to be determined dynamically. In contrast to ref.\ \cite{Kovensky:2023mye} (and in agreement with ref.\ \cite{Kovensky:2021ddl}), our convention for the isospin chemical potential is such that the zero-temperature onset of pion condensation is at $\mu_I=m_\pi/2$, corresponding to $\mu_I=(\mu_{u}-\mu_d)/2$ in terms of quark chemical potentials. The angles in eq.\ (\ref{K123inf}) are the ones from the chiral field (\ref{angles}). In the rotated frame, all fields are either even or odd in $z$ and we can restrict ourselves to one half of the connected flavor branes, working with the holographic coordinate  $u$ (the factor 2 from the two halves is already taken into account in the prefactor of the DBI action). 

%%%%%%%%%%%%%%%%%%%%%%%%%%%%%%%%%%%%%%%%%%%%
\subsubsection{Equations of motion}
%%%%%%%%%%%%%%%%%%%%%%%%%%%%%%%%%%%%%%%%%%%%

We define an effective dimensionless potential, obtained from the DBI action and the mass term,
\be
\Omega = \frac{T}{V}\frac{S}{{\cal N}N_f} = \int_{u_c}^\infty du\, u^{5/2}\sqrt{1+u^3f_Tx_4'^2-K_a'K_a'} -\frac{A\cos\theta}{2\lambda_0} \, , 
\ee
where 
\be \label{A0}
A = \frac{2\alpha}{\lambda_0^2}\exp\left(2\lambda_0\int_{u_c}^\infty du\, \phi_T x_4'\right) \, .
\ee
The integrated equations of motion for $K_a$ and $x_4$ are
\begin{subequations}
\bea
\frac{u^{5/2}K_a'}{\sqrt{1+u^3f_Tx_4'^2-K_b'K_b'}}&=&\kappa_a \, , \label{eomK} \\[2ex]
\frac{u^{11/2}f_Tx_4'}{\sqrt{1+u^3f_Tx_4'^2-K_b'K_b'}} &=& k+A\phi_T\cos\theta \, , 
\eea
\end{subequations}
with integration constants $\kappa_a$, $k$. These equations can be algebraically solved for the derivatives of the fields, 
\begin{subequations} \label{sol}
\bea
K_a' &=& \frac{\kappa_a}{u^{5/2}}\zeta \,, \label{solKa} \\[2ex]
x_4' &=& \frac{k+A\phi_T\cos\theta}{u^{11/2}f_T}\zeta \, ,
\eea
\end{subequations}
and $\zeta$ assumes the expression 
\be
\zeta = \left[1-\frac{(k+A\phi_T\cos\theta)^2}{u^8f_T}+\frac{\kappa^2}{u^5}\right]^{-1/2} \, ,
\ee
where $\kappa^2 = \kappa_a \kappa_a$.
With $\zeta (u=\infty) = 1$ we can extract the asymptotic behavior of the gauge fields, $(u^{5/2}K_a')_{\infty} = \kappa_a $, which, according to the AdS/CFT dictionary should give the density associated with the chemical potential at the boundary. Indeed, we shall see below that the integration constants $\kappa_a$ capture the different contributions to the isospin density. 

%%%%%%%%%%%%%%%%%%%%%%%%%%%%%%%%%%%%%%%%%%%%
\subsubsection{Stationarity of the effective potential}
%%%%%%%%%%%%%%%%%%%%%%%%%%%%%%%%%%%%%%%%%%%%

Besides the integration constants $k$, $\kappa_a$, and the yet unknown boundary value $K_3(u_c)$, the effective potential also contains the parameters $u_c$ and $\theta$. To compute all these unknowns we first bring the effective potential into a convenient form. Using eq.~\eqref{ell} together with 
\be \label{uKK}
\int_{u_c}^\infty du\, \frac{u^{5/2}K_a'K_a'}{\zeta} 
%=\int_{u_c}^\infty du\, \der_u \left(u^{5/2}\zeta^{-1}K_a'K_a\right) 
=\kappa_a [K_a(\infty)-K_a(u_c)] \, ,
\ee
where partial integration has been employed, we compute 
\bea \label{Omega0}
\Omega &=& 
%\int_{u_c}^\infty du\, u^{5/2}\zeta - \frac{A\cos\theta}{2\lambda_0} =
\int_{u_c}^\infty du\left(\frac{u^{5/2}}{\zeta}+A\phi_Tx_4'\cos\theta\right)-\frac{A\cos\theta}{2\lambda_0} +k \, \frac{\ell}{2} \non[2ex]
&&  -\bar{\mu}_I\Big[(-\kappa_1\sin\varphi+\kappa_2\cos\varphi)\sin\theta+\kappa_3\cos\theta\Big]+\kappa_3K_{3}(u_c) \, . 
\eea
The free parameters can now be straightforwardly computed by requiring the potential to be stationary with respect to them. First, 
we find
\be
0 = \frac{\partial \Omega}{\partial u_c} = - \frac{u_c^{5/2}}{\zeta(u_c)} \, , 
\ee
which can be solved for $k$,   
\be
k = u_c^4\sqrt{f_T(u_c)\left(1+\frac{\kappa^2}{u_c^5}\right)}-A\phi_T(u_c)\cos\theta \, .
\ee
This condition is equivalent to the smoothness of the flavor brane embedding at the tip of the connected branes, $x_4'(u\to u_c)=\infty$.  Next, stationarity with respect to $k$ turns out to be equivalent to the boundary condition \eqref{ell} and thus does not yield any new constraint. Stationarity with respect to $\kappa_a$ can be written as 
\begin{subequations} \label{dk123}
\bea
0 &=& \frac{\partial \Omega}{\partial \kappa_1} =
\kappa_1 \int_{u_c}^\infty du\, \frac{\zeta}{u^{5/2}}+\bar{\mu}_I\sin\theta\sin\varphi \, , \label{dk1}\\[2ex]
0 &=& \frac{\partial \Omega}{\partial \kappa_2} =
\kappa_2 \int_{u_c}^\infty du\, \frac{\zeta}{u^{5/2}}-\bar{\mu}_I\sin\theta\cos\varphi \, , \label{dk2}\\[2ex]
0 &=& \frac{\partial \Omega}{\partial \kappa_3} =
\kappa_3 \int_{u_c}^\infty du\, \frac{\zeta}{u^{5/2}}-\bar{\mu}_I\cos\theta+K_{3}(u_c) \label{dk3}\, . 
\eea
\end{subequations}
These equations can also be obtained more directly by inserting eq.\ (\ref{solKa}) into eq.\ (\ref{uKK}) [or simply by integrating eq.\ (\ref{solKa})].  
Stationarity with respect to $K_3(u_c)$ yields 
\bea \label{kap30}
0 &=& \frac{\partial \Omega}{\partial K_3(u_c)} = \kappa_3 \, ,
\eea
which, combined with \eqref{dk3}, implies  
\begin{equation}
    K_{3}(u_c) = \bar{\mu}_I \cos\theta \, ,
\end{equation}
which shows that ultraviolet and infrared boundary values of $K_3$ are the same. In fact, from eqs.\ (\ref{solKa}) and (\ref{kap30}) we conclude that $K_3(u)$ is constant in $u$. Therefore, it does not contribute to the isospin density. 
The isospin density is entirely given by $\kappa_1$ and $\kappa_2$. More precisely, only the modulus $\kappa^2 = \kappa_1^2+\kappa_2^2$ is relevant because the angle $\varphi$ should play no role in any physical observable. This becomes manifest by combining eqs.~(\ref{dk1}) and (\ref{dk2}) to
\be \label{kapkap}
\kappa \int_{u_c}^\infty du\, \frac{\zeta}{u^{5/2}}= \bar{\mu}_I\sin\theta \, .
\ee
Using eqs.\ (\ref{Omega0}), (\ref{dk1}), (\ref{dk2}), we can compute the isospin density from its thermodynamic definition,
\be
\label{defnIpi}
\bar{n}_I = -\frac{\partial \Omega}{\partial \bar{\mu}_I} 
= \left(\int_{u_c}^\infty du\, \frac{\zeta}{u^{5/2}}\right)^{-1} \bar{\mu}_I \sin^2\theta \, .
\ee
With eqs.\ (\ref{kapkap}) and (\ref{defnIpi}) we have $\kappa=\bar{n}_I/\sin\theta$, which can be used to eliminate $\kappa$ from the calculation. It is instructive to write the isospin density in terms of
dimensionful quantities. Employing eqs.\ (\ref{normN}), (\ref{muIdim}), (\ref{nIq}), we write eq.\ (\ref{defnIpi}) in the form 
\be
n_I = 4F_\pi^2 \mu_I\sin^2\theta \, , 
\ee
with the pion decay constant defined in eq.~(\ref{Fpim}). This is exactly the form of leading-order chiral perturbation theory, but contains the effect of the medium in $F_\pi$.  

Finally, we need to minimize the effective potential with respect to $\theta$,
\be\label{theta}
0 = \frac{\partial \Omega}{\partial \theta} =  \sin\theta\left(\frac{A}{2\lambda_0} - \bar{\mu}_I\bar{n}_I\frac{\cos\theta}{\sin^2\theta}\right) \, .
\ee
This equation is trivially fulfilled for vanishing pion condensate $\theta=0$. We also see from this condition that, assuming $\kappa\neq 0$,  the condensate can only be zero or maximal ($\theta=\pi/2$) if $A=0$, which corresponds to the chiral limit $m_\pi=0$. For nonzero pion mass the condensate can assume nontrivial values, which depend on temperature and isospin chemical potential.
Furthermore, eq.\ (\ref{theta}) can be used to derive a condition for the onset of pion condensation. After dividing by $\sin\theta$ we take the limit $\theta\to 0$ to derive the onset chemical potential 
\be \label{muIc}
\bar{\mu}_{I,{\rm onset}}^2 = \frac{A}{2\lambda_0} \int_{u_c}^\infty du\frac{\zeta}{u^{5/2}} \,,
\ee
where $A$, $\zeta$, and $u_c$ are evaluated at vanishing pion condensate. This equation is useful because it gives an expression of the onset that can be computed by solely working within the isospin vacuum. For a more physical form of this condition, we use eqs.\ (\ref{qqc}), (\ref{qqc0}), (\ref{A}), (\ref{alphadef}), and (\ref{Fpim}) to obtain
\be
\mu_{I,{\rm onset}} = \frac{m_\pi}{2}\frac{f_\pi}{F_\pi}\left(\frac{\langle\bar{\psi}\psi\rangle}{\langle\bar{\psi}\psi\rangle_0}\right)^{1/2} \, .
\ee
Since the isospin vacuum does not depend on $\mu_I$, the right-hand side is a function of $T$ only (in general, $F_\pi$ and $\langle\bar{\psi}\psi\rangle$ are functions of $\mu_I$ and $T$, here they are understood to be evaluated at $\mu_I$=0). We see that as $T\to 0$, the onset chemical potential reduces to $\mu_{I,{\rm onset}}=m_\pi/2$, in agreement with zero-temperature $\chi$PT.

To summarize, for given $\bar{\mu}_I$ and $T$ we need to solve the coupled equations  (\ref{ell}), (\ref{A0}), (\ref{defnIpi}), (\ref{theta}) for the variables $A$, $\bar{n}_I$, $u_c$,  $\theta$. This has to be done numerically. The solution is then inserted back into the effective potential to obtain the free energy, 
\be\label{Omegas}
\Omega =  \int_{u_c}^\infty du\, u^{5/2}(\zeta-1) -\frac{2}{7}u_c^{7/2}- \frac{A\cos\theta}{2\lambda_0}  + \frac{A_0}{2\lambda_0} \, .
\ee
Here we have subtracted the (infinite) vacuum contribution, with $A_0\equiv A(T=\mu_I=0)$ at fixed nonzero pion mass. The integral that is left is finite.  

%%%%%%%%%%%%%%%%%%%%%%%%%%%%%%%%%%%%%%%%%%%%
\subsubsection{Speed of sound}
\label{sec:sound}
%%%%%%%%%%%%%%%%%%%%%%%%%%%%%%%%%%%%%%%%%%%%

We will also be interested in computing the speed of sound. To this end, we need the (dimensionless) entropy density, 
\bea
s &=& -\frac{\partial\Omega}{\partial t} = -2\frac{u_T}{t}\frac{\partial\Omega}{\partial u_T}  \, .
\eea
Starting from the free energy in the form \eqref{Omega0}, we compute
\bea
s &=& \frac{2}{t}\int_{u_c}^\infty du\, x_4'(u)\left[\frac{3u_T^3}{2u^3}\frac{k+A\phi_T(u)\cos\theta}{f_T(u)}+A\cos\theta\left(\phi_T(u)-\frac{u}{\sqrt{f_T(u)}}\right)\right] \label{entropy pi-phase} \, , 
\eea
where we have used 
\be
\frac{\partial \phi_T}{\partial u_T} = \frac{1}{u_T}\left[\phi_T(u)-\frac{u}{\sqrt{f_T(u)}}\right] \, .
\ee
We then compute the dimensionless pressure and energy density, using the normalized free energy (\ref{Omegas}),
\be
P = -\Omega \, , \qquad \epsilon = \Omega + \bar{\mu}_I\bar{n}_I +ts \, ,
\ee
and  calculate the speed of sound via 
\be
c_s^2 = \left.\frac{\partial P}{\partial \epsilon}\right|_{s/\bar{n}_I} \, .
\ee
We take this derivative numerically by solving the equations of motion at two different points, for instance distinguished by two different densities $\bar{n}_I$ and $\bar{n}_I+\delta\bar{n}_I$, where $\delta\bar{n}_I$ is sufficiently small and both points have the same entropy per particle $s/\bar{n}_I$.

%%%%%%%%%%%%%%%%%%%%%%%%%%%%

%%%%%%%%%%%%%%%%%%%%%%%%%%%%%%%%%%%%%%%%%%%%%%%%%%%%%%%%%
\section{Parameter fit}
\label{sec:fit}
%%%%%%%%%%%%%%%%%%%%%%%%%%%%%%%%%%%%%%%%%%%%%%%%%%%%%%%%%

In this section we discuss our choice of the model parameters, which will be used for deriving the results presented in section \ref{sec:results}. With $N_c=3$, the only model parameters are $L$, $\lambda$, and $m_\pi$. Equivalently, and more conveniently, we will work instead with $L$, $\tilde{\lambda}_0\equiv \lambda_0/\ell$, and $\tilde{\alpha}=\ell^4\alpha$. 
We will choose these three parameters such that we reproduce the physical pion mass, pion decay constant, and the critical temperature of the chiral transition at $\mu_I=0$. The procedure for this fit is as follows. 

In the dimensionless form in which the stationarity and self-consistency equations of all our phases are written, the only model parameters that appear are $\ell$ and $\alpha$. Following earlier works in the deconfined, decompactified limit of the WSS model \cite{Li:2015uea,Kovensky:2019bih,Kovensky:2020xif}, we observe that $\ell$ can be eliminated from these equations by a rescaling with a suitable power of $\ell$ of all variables and the holographic coordinate. Denoting the rescaled quantities by a tilde, we have
$\tilde{A} = \ell^6 A$, $\tilde{u}=\ell^2 u$, $\tilde{n}_I=\ell^5\bar{n}_I$, $\tilde{\mu}_I = \ell^2\bar{\mu}_I$, $\tilde{t} = \ell t$. In the vacuum, we can write the pion decay constant, see eqs.\ (\ref{Fpim}) and (\ref{fpiFpi}), as
\bea
\label{fpi}
f_\pi^2 &=& 
\frac{\tilde{\lambda}_0}{4\pi^2 L^2}
\left(\int_{\tilde{u}_c}^\infty d\tilde{u} \frac{\zeta}{\tilde{u}^{5/2}}\right)^{-1} \, , 
\eea
and we  use the definition of $\alpha$ (\ref{alphadef})
in the form 
\bea
\tilde{\alpha} &=& \frac{\pi^2f_\pi^2m_\pi^2L^4}{e^{-S_{\rm NG}^{0}}} \, .\label{alL}
\eea
To determine the model parameters, these equations must be coupled with the equations for the vacuum (\ref{ell2Mes}),
\begin{subequations}\label{vacfit}
\bea
\frac{1}{2}&=& \int_{\tilde{u}_c}^\infty d\tilde{u}\,\tilde{x}_{4}'  \,, \label{vacfit1}\\[2ex]
\tilde{A} &=& \frac{m_\pi^2}{8\pi^2 f_\pi^2}\left(\int_{\tilde{u}_c}^\infty d\tilde{u}\frac{\zeta}{\tilde{u}^{5/2}}\right)^{-2}\label{tildeA} 
\, .
\eea
\end{subequations}
Here we have used eqs.\ (\ref{fpi}) and (\ref{alL}) to rewrite eq.\ (\ref{AIvac}) in the form (\ref{tildeA}). The advantage is that now the system of equations (\ref{vacfit}) does not depend on any model parameters and can be solved numerically for $\tilde{A}$ and $\tilde{u}_c$ for given $f_\pi$ and $m_\pi$. The results give $\tilde{\alpha}$ (\ref{alL}) and $\tilde{\lambda}_0$ as functions of $L$, where the latter can for instance be obtained by combining eqs.\ (\ref{A}) and (\ref{alphadef}), 
\bea
\tilde{\lambda}_0 &=& \sqrt{\frac{2}{\tilde{A}}}\, \pi f_\pi m_\pi L^2 \, .
\eea
It remains to fix the length scale $L$. While reproducing the physical values of $f_\pi$ and $m_\pi$ seems very natural in our context, there are different possibilities to choose the third physical quantity for our parameter fit. Our choice is motivated by the $T$-$\mu_I$ phase diagram: since $m_\pi$ determines the phase transition on the $\mu_I$ axis, we decide to reproduce the location of the real-world phase transition on the $T$ axis as well. Then, our phase structure is ``anchored'' by these two points. As we shall see, and as is already known from previous studies, the phase transition at vanishing chemical potentials in the WSS model is of first order. In the real world, however, it is a crossover, and thus we are matching the critical temperature of our holographic first-order transition 
(= coexistence of isospin vacuum and HTQ phase at $\mu_I=0$) to the pseudo-critical temperature of QCD. 

With the help of eq.\ (\ref{tdef}) we write the critical temperature  
as
\begin{equation}
\label{Tc 160 MeV condition} 
T_c = \frac{3 \sqrt{\tilde{u}_{T_c}}}{4\pi L} \, .
\end{equation}
This equation has to be coupled to the equations that determine the critical temperature in our model. These are the two equations (\ref{ell2Mes}) for the isospin vacuum, eq.\ (\ref{AHTQ}) for the HTQ phase at $n_I=0$, and the condition that the two free energies (\ref{OM IVac}) and (\ref{OM HTQ}) be the same. Inserting $\tilde{\lambda}_0$ and $\tilde{\alpha}$ from above, these are 5 equations in total to be solved for $\tilde{u}_c$ and $\tilde{A}$ of the isospin vacuum, $\tilde{A}$ of the HTQ phase, as well as $\tilde{u}_{T_c}$ and $L$. For a given numerical value of $T_c$, this fixes the remaining parameter $L$.

\begin{figure} [t]
\begin{center}
\hbox{
\includegraphics[width=0.5\textwidth]{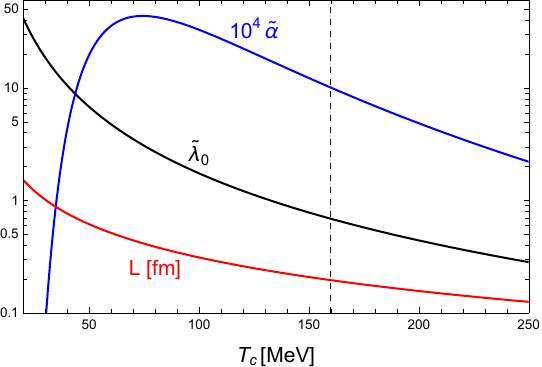}
\includegraphics[width=0.48\textwidth]{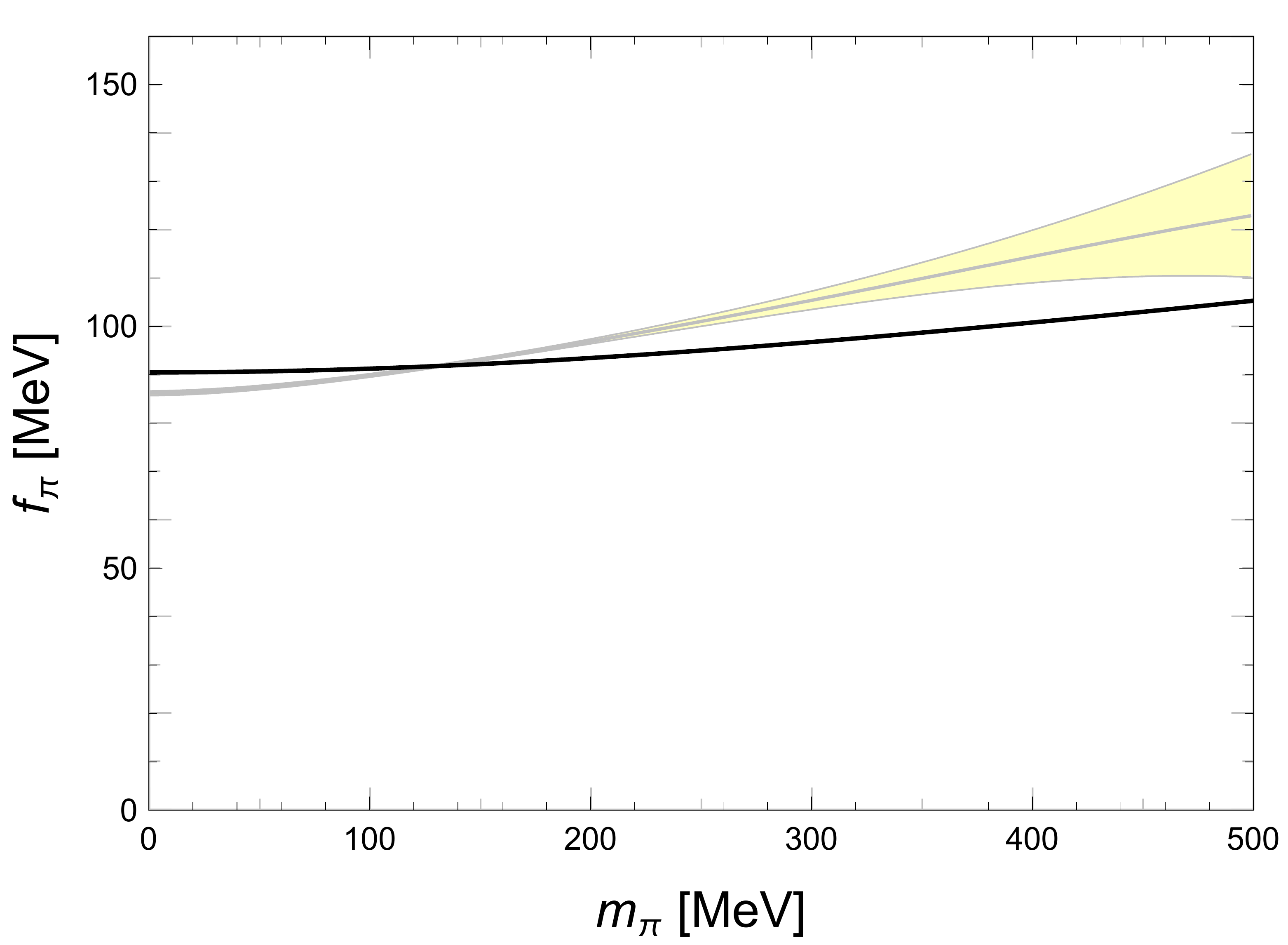}}
\caption{\textit{Left panel:}
Model parameters $\tilde{\alpha}$, $\tilde{\lambda}_0$, $L$ for fixed  $m_\pi=140\, {\rm MeV}$, $f_\pi=92\, {\rm MeV}$ as a function of the critical temperature $T_c$ at $\mu_I=0$. The main results in section \ref{sec:results} are obtained by choosing the physical value $T_c=160\, {\rm MeV}$ (dashed vertical line). 
\textit{Right panel:} Pion decay constant as a function of the pion mass, with $\tilde{\lambda}_0$ and $L$ fixed as in eq.\ (\ref{Lfit}), compared to the result from next-to-next-to-leading-order $\chi$PT of ref.\ \cite{Colangelo:2003hf}. }
\label{fig:para0}
\end{center}
\end{figure}

With $m_\pi=140\, {\rm MeV}$, $f_\pi = 92\, {\rm MeV}$, $T_c=160\, {\rm MeV}$, this calculation yields 
\be
\label{Lfit}
\tilde{\lambda}_0 \simeq 0.6856 \, , \qquad \tilde{\alpha} \simeq 1.005\times 10^{-3} \, , \qquad 
L \simeq  
%\left( 1012\, {\rm MeV}\right)^{-1} \simeq 
0.1950\, {\rm fm} \, .
\ee
Since using the critical temperature for our fit is somewhat arbitrary, we show the dependence of our model parameters as a function of $T_c$ in the left panel of figure \ref{fig:para0}. 

It is instructive to compare our parameter set \eqref{Lfit} with previous parameter choices in the WSS model. The 't Hooft coupling can be inferred from our numerical values via $\lambda =4\pi\tilde{\lambda}_0LM_{\rm KK}$, which requires a choice of the Kaluza-Klein scale. The fit to rho meson mass and pion decay constant used in the original works \cite{Sakai:2004cn,Sakai:2005yt} gives $\lambda \simeq 16$ and $\MKK \simeq 949$ MeV (however, in the confined geometry, i.e., this comparison is mainly of illustrative nature). Using this value of $\MKK$, our fit yields the smaller value $\lambda \simeq 8.1$. We also get a smaller value of $\tilde{\lambda} = 4\pi\tilde{\lambda}_0 \simeq 8.6$ compared to the somewhat arbitrarily chosen $\tilde{\lambda}=15$ of refs.\
\cite{Kovensky:2019bih,Kovensky:2020xif} and $\tilde{\lambda}=20,40$ of ref.\ \cite{Kovensky:2021ddl} (which all consider the same decompactified limit as the present work). Interestingly, our value for $\lambda$ is closer to the one that reproduces properties of nuclear matter at saturation, as discussed in the context of holographic neutron stars \cite{Kovensky:2021kzl, Kovensky:2021wzu,Kovensky:2023mye}. Our value for the ``mass parameter'' $\tilde{\alpha}$ is similar to the physical value identified in refs.\ \cite{Kovensky:2019bih,Kovensky:2020xif}, where, however,  $\tilde{\alpha}$ was varied as a free parameter to explore ``heavy holographic QCD''.  

To get an idea of the $m_\pi$ dependence of our results we keep the model parameters $\tilde{\lambda}_0$ and $L$ fixed as in eq.\ (\ref{Lfit}) and re-calculate $\tilde{\alpha}$ as a function of $m_\pi$. (As a consequence, $f_\pi$ and $T_c$ will become functions of $m_\pi$ as well.)  To this end, we insert eq.\ (\ref{fpi}) into eq.\ (\ref{tildeA}) to obtain 
\bea
\frac{2\tilde{\lambda}_0\tilde{A}}{m_\pi^2L^2} &=& \left(\int_{\tilde{u}_c}^\infty d\tilde{u}\frac{\zeta_{0}}{\tilde{u}^{5/2}}\right)^{-1} \, .
\eea
This equation can now be solved together with eq.\ (\ref{vacfit1}) for $\tilde{A}$ and $\tilde{u}_c$ without having to make a choice for $f_\pi$. Then, the $m_\pi$-dependent pion decay constant is 
\be
f_\pi^2 = \frac{\tilde{\lambda}_0^2\tilde{A}}{2\pi^2m_\pi^2L^4} \, , 
\ee
and the corresponding $\tilde{\alpha}$ can be computed from eq.\ (\ref{alL}). We show the pion decay constant as a function of $m_\pi$ in the right panel of figure \ref{fig:para0} and compare it to the result obtained in next-to-next-to-leading-order $\chi$PT. We see that our mass dependence is milder than from that calculation. The two results coincide at the physical point, as it should be by construction. 

In section \ref{sec:results} we shall compare our results to some lattice results obtained at $m_\pi=170\, {\rm MeV}$. For a meaningful comparison, we use the calculation just explained to adjust our model parameters to this case and find $\tilde{\alpha}\simeq 1.496\times 10^{-3}$, with a pion decay constant $f_\pi\simeq 92.68\, {\rm MeV}$. 
This comparison will not show any inconsistencies caused by the mildness of the $m_\pi$ dependence in our pion decay constant compared to $\chi$PT. Therefore, this discrepancy -- possibly related to the fact that our effective mass term is constructed only from the lowest-order mass correction -- is of minor significance for our main conclusions. 
We will also vary the pion mass at the end of section \ref{sec:results} in the phase diagram of figure \ref{fig:phasediagram}, where the pion masses $m_\pi = (10,70,280)\, {\rm MeV}$ require the use of the parameters $\tilde{\alpha} = (5.018\times 10^{-6},2.472\times 10^{-4},4.243\times 10^{-3})$, together with the fixed values of $\lambda_0$ and $L$.

%%%%%%%%%%%%%%%%%%%%%%%%%%%%%%%%%%%%%%%%%%%%%%%%%%%%%%%%%
\section{Results}
\label{sec:results}
%%%%%%%%%%%%%%%%%%%%%%%%%%%%%%%%%%%%%%%%%%%%%%%%%%%%%%%%%

We are now prepared for the numerical evaluation. For given model parameters (\ref{Lfit}) and values of the thermodynamic variables $\mu_I$ and $T$, we solve the relevant equations for each phase numerically and insert the results back into the effective potential to obtain the free energy. We can thus determine the energetically preferred phase for any $(\mu_I,T)$ and compute the resulting phase transitions. In the following, when we present various thermodynamic observables we always show the corresponding value  in the favored phase. 

%%%%%%%%%%%%%%%%%%%%%%%%%%%%%%%%%%%%%%%%%%%%%%%%%%%%%%%%%%%%%%%%%
\subsection{Zero temperature}
\label{sec:results-zeroT}
%%%%%%%%%%%%%%%%%%%%%%%%%%%%%%%%%%%%%%%%%%%%%%%%%

\begin{figure} [t]
\begin{center}
\includegraphics[width=0.52\textwidth]{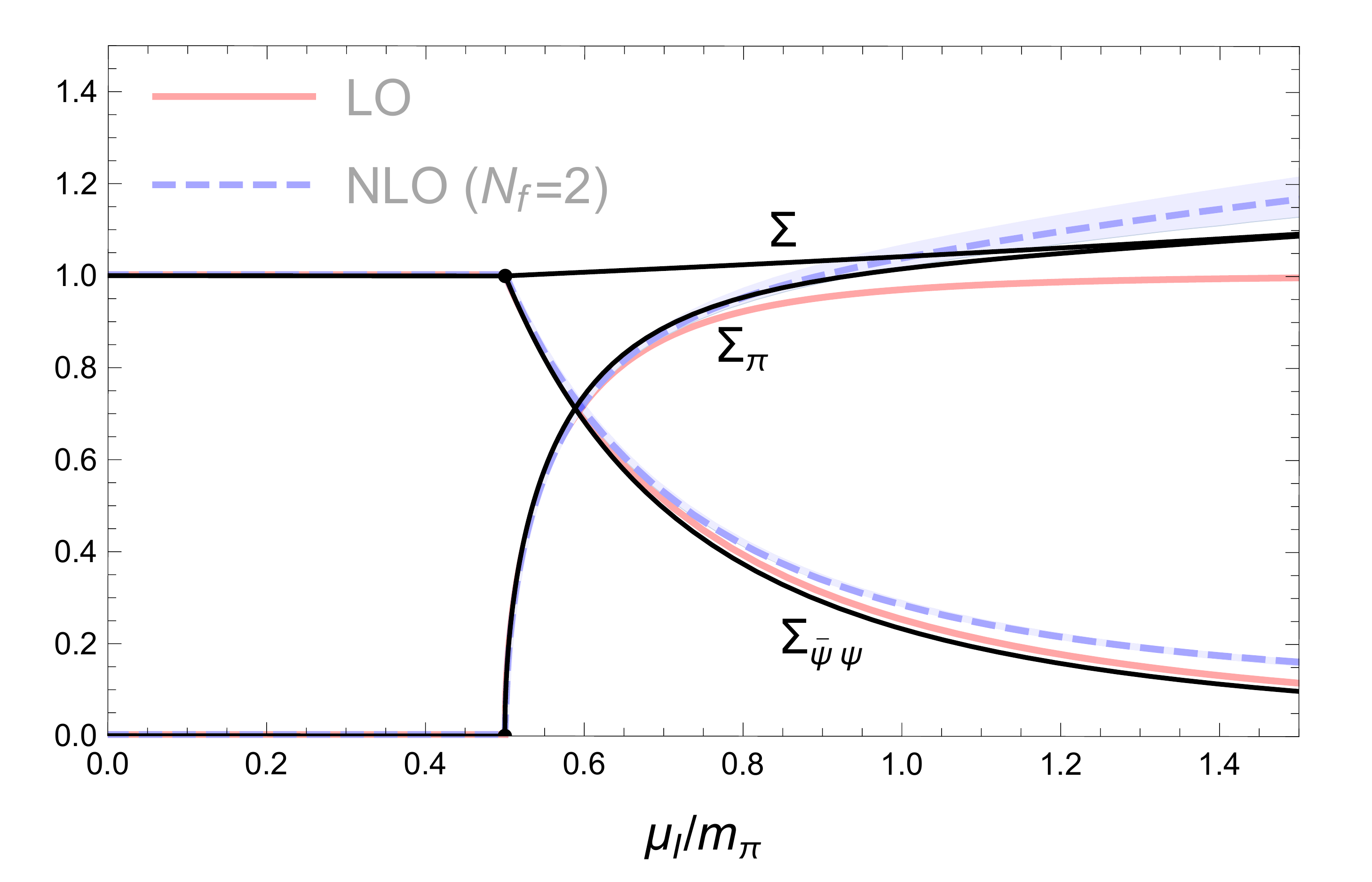}
\caption{Normalized pion and chiral condensates $\Sigma_\pi$ and $\Sigma_{\bar{\psi}\psi}$, defined in eq.\ (\ref{SigSig}), as a function of $\mu_I$ at $T=0$, compared to the leading-order (LO) and next-to-leading order (NLO) $\chi$PT results from ref.\ \cite{Adhikari:2020ufo}. The ``total condensate'' $\Sigma$ deviates from 1 in the presence of pion condensation, as expected from $\chi$PT beyond leading order. } 
\label{fig:SigSigmu}
\end{center}
\end{figure}

We first focus on zero temperature and discuss the behavior of our system as a function of $\mu_I$. It is useful to define the normalized chiral and pion condensates by dividing those in eq.\ (\ref{psipsipi}) by the chiral condensate in the vacuum (at the physical pion mass),
\begin{subequations}\label{SigSig}
\bea
\Sigma_{\bar{\psi}\psi} \equiv \frac{\langle \bar{\psi}\psi\rangle}{\langle \bar{\psi}\psi\rangle_0} &=& \frac{{\cal N}A}{\lambda_0 m_\pi^2f_\pi^2}\cos\theta \, , \\[2ex] 
\Sigma_{\pi} \equiv \frac{\langle \pi^\pm \rangle}{\langle \bar{\psi}\psi\rangle_0} &=& \frac{{\cal N}A}{\lambda_0 m_\pi^2f_\pi^2}\sin\theta \, .
\eea
\end{subequations}
In leading-order $\chi$PT, these two condensates form a vector with unit length, $\Sigma_{\bar{\psi}\psi}^2+\Sigma_{\pi}^2=1$. This is not true in general, as results from $\chi$PT beyond leading order \cite{Adhikari:2020ufo} and on the lattice \cite{Brandt:2022hwy} show. 
Our holographic model also violates this constraint, 
\be\label{constraint}
\Sigma^2 \equiv \Sigma_{\bar{\psi}\psi}^2+\Sigma_{\pi}^2 = \left(\frac{{\cal N}A}{\lambda_0 m_\pi^2f_\pi^2}\right)^2 \, .
\ee
In figure \ref{fig:SigSigmu} we show both condensates and the ``total condensate'' $\Sigma$ as a function of $\mu_I$. As expected from the results in section \ref{sec:pi-phase}, we find that the onset of pion  condensation is at $\mu_I=m_\pi/2$, the free energy of this phase is indeed lower than that of the isospin vacuum as soon as it exists. Close to the onset, our results agree with leading-order $\chi$PT, while they deviate for larger values of $\mu_I$. This is expected due to the non-antipodal separation of the flavor branes and the resulting backreaction of the pion condensate on the embedding function. Our results agree with next-to-leading order $\chi$PT qualitatively in the sense that $\Sigma_\pi$ and thus also $\Sigma$ can become greater than 1. However, our holographic  calculation is obviously not performed at a fixed order of $\chi$PT and thus it is no surprise that, quantitatively, we do not obtain a precise match with next-to-leading order $\chi$PT. 

\begin{figure} [t]
\begin{center}
\hbox{
\includegraphics[width=0.47\textwidth]{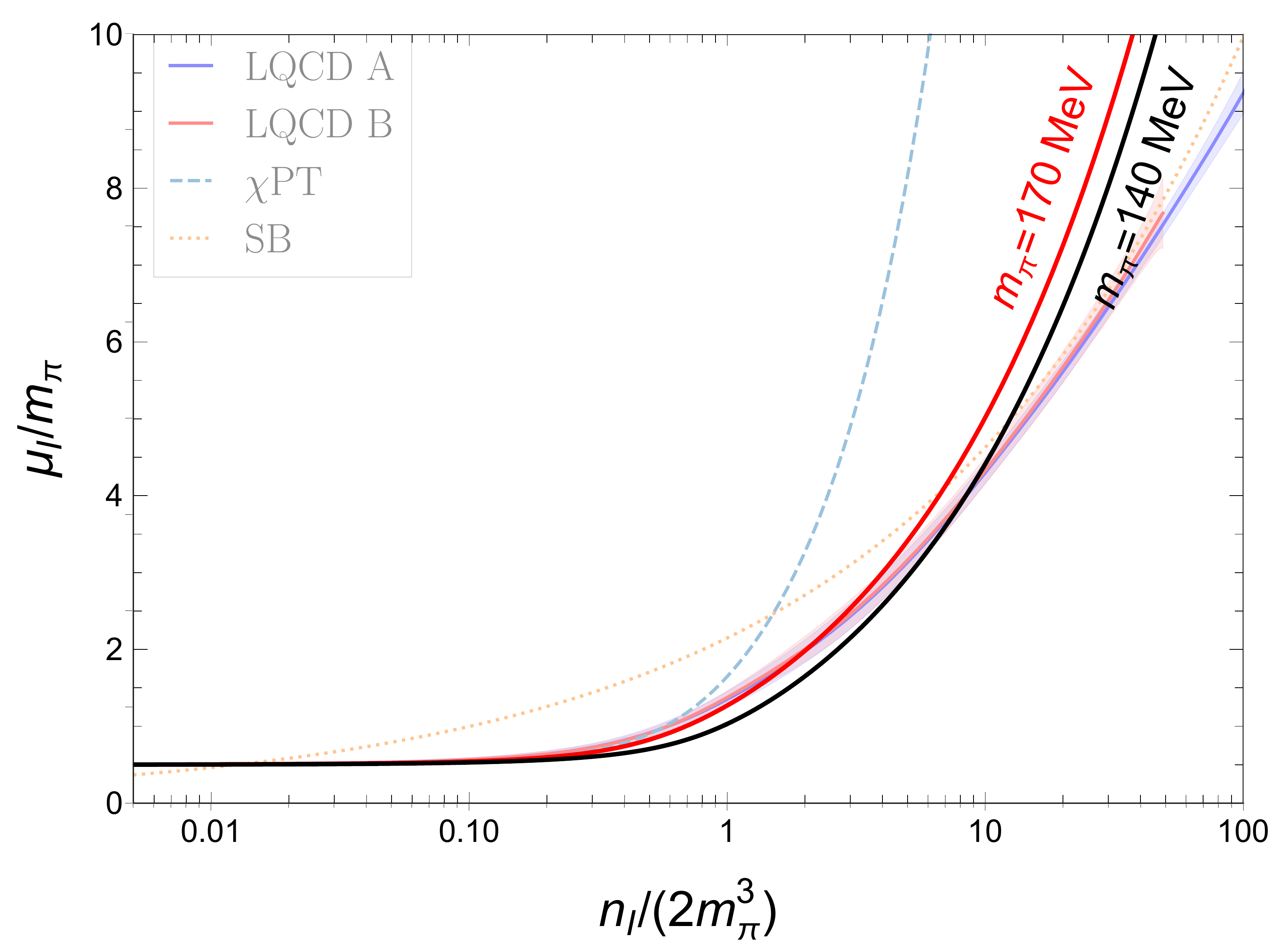}
\includegraphics[width=0.5\textwidth]{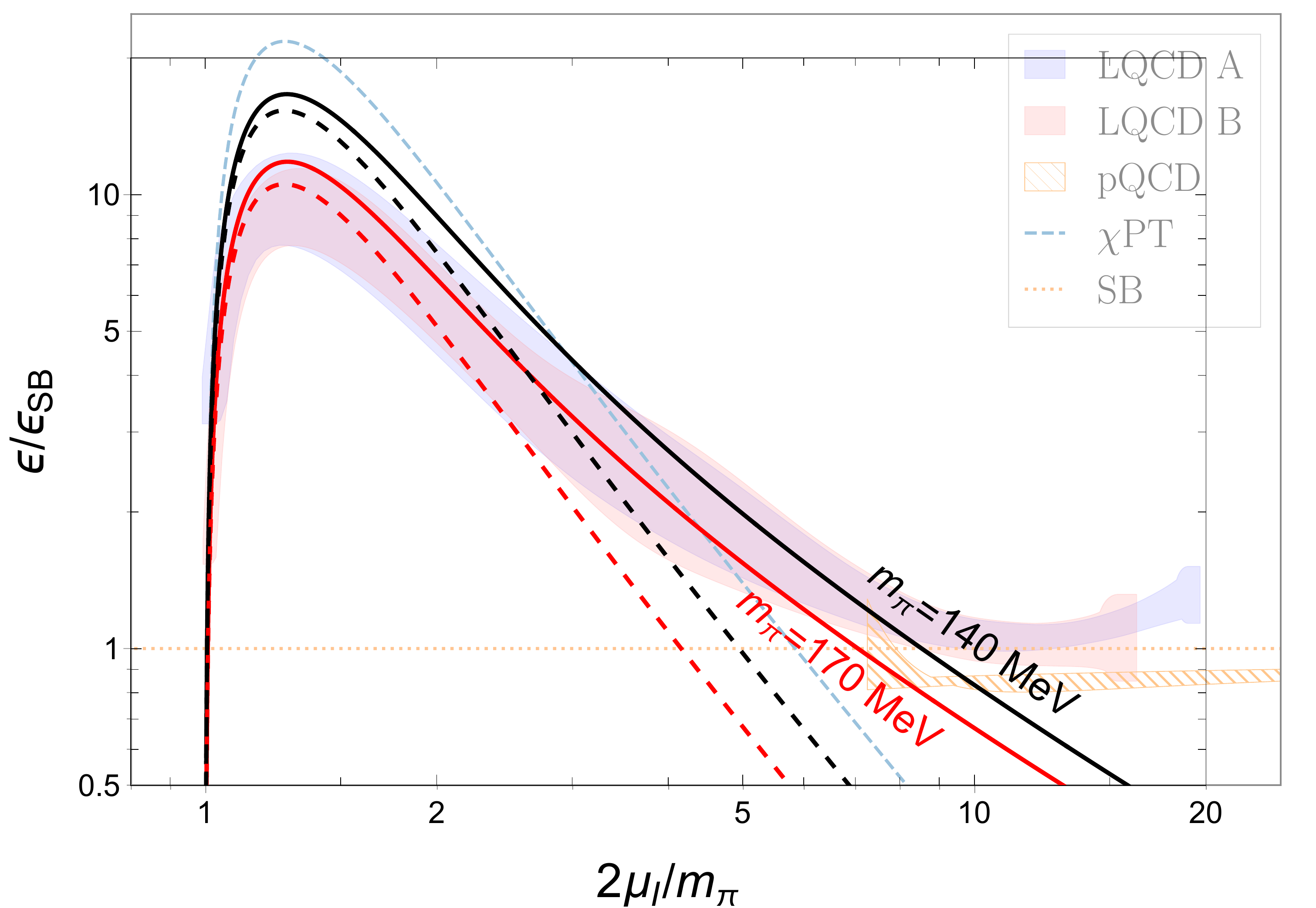}}
\caption{{\it Left panel:} Isospin chemical potential as a function of (half of the) isospin density at $T=0$ for two values of the pion mass, compared to the lattice results of ref.\ \cite{Abbott:2023coj}, where $m_\pi=170\, {\rm MeV}$ was used
(red and blue bands for two different lattice configurations).
{\it Right panel:} Energy density normalized by the 
Stefan-Boltzmann limit as a function of (twice the) isospin chemical potential, for the same two pion masses and compared 
to the same lattice calculation as in the left panel, with  perturbative data from ref.\ \cite{Graf:2015tda} added. The dashed red and black lines are the corresponding leading-order 
$\chi$PT results.}
\label{fig:nIeps}
%\vspace{-0.9cm}
\end{center}
\end{figure}

It is therefore interesting to compare our results to lattice QCD data. In figure \ref{fig:nIeps} we superpose our results for the isospin density and the energy density with the results of ref.\ \cite{Abbott:2023coj}\footnote{In the right panel, the (blue dashed) $\chi$PT curve of ref.\ \cite{Abbott:2023coj} is incorrect by a factor 2, hence the discrepancy to our red dashed curve; we thank Ryan Abbott, co-author of ref.\ \cite{Abbott:2023coj}, for confirming this error, which does not affect the lattice results.}.
(See ref.\ \cite{Chiba:2023ftg} for a comparison of the quark-meson model with the same lattice results.) This reference uses a convention of the isospin chemical, and hence also the isospin density, which differs by a factor 2 from ours. We have taken this into account by using half of our isospin density $n_I$ and twice our chemical potential $\mu_I$ in the horizontal logarithmic scales. Moreover, in the right panel, we have normalized the energy density by that of a free Fermi gas with $N_f=2$, $N_c=3$, the Stefan-Boltzmann limit $\ep_{\rm SB} = 3\mu_I^4/(2\pi^2)$. We show the results for both $m_\pi=170\, {\rm MeV}$, which was used in ref.\ \cite{Abbott:2023coj}, and the physical pion mass $m_\pi=140\, {\rm MeV}$. The former requires an adjustment of our model parameters as explained at the end of section \ref{sec:fit}. (This adjustment results in a slightly different value for the pion decay constant compared to that used in ref.\ \cite{Abbott:2023coj}, which however is negligible for our purpose.)  

Remarkably, isospin density and energy density are in excellent agreement with the lattice results  for isospin chemical potentials up to $\mu_I \sim 3m_\pi$. In the region where our results deviate, the lattice results are already close to being well approximated by the Stefan-Boltzmann limit. This is an indication that, at least with respect to the observables considered in this figure, QCD already behaves similar to a  weakly coupled system. It is therefore not surprising that our holographic results are no longer valid in this regime; our classical gravity approximation in the bulk is expected to work well in the strong-coupling regime of the dual field theory and in particular does not exhibit asymptotic freedom. We should also mention that we have not taken into account the possibility of rho meson condensation. Previous works in the WSS model -- albeit in the confined, antipodal version -- predict condensation to set in at around $\mu_I \simeq 4.7\, m_\pi$ \cite{Aharony:2007uu,Kovensky:2023mye} (at $1.7m_\rho\simeq 9.3\, m_\pi$ in these works due to the difference in the convention for the isospin chemical potential). It is therefore conceivable that our results at these very large isospin densities can be further improved by allowing for a more general ansatz for the gauge fields in the bulk.

\begin{figure} [t]
\begin{center}
\includegraphics[width=0.5\textwidth]{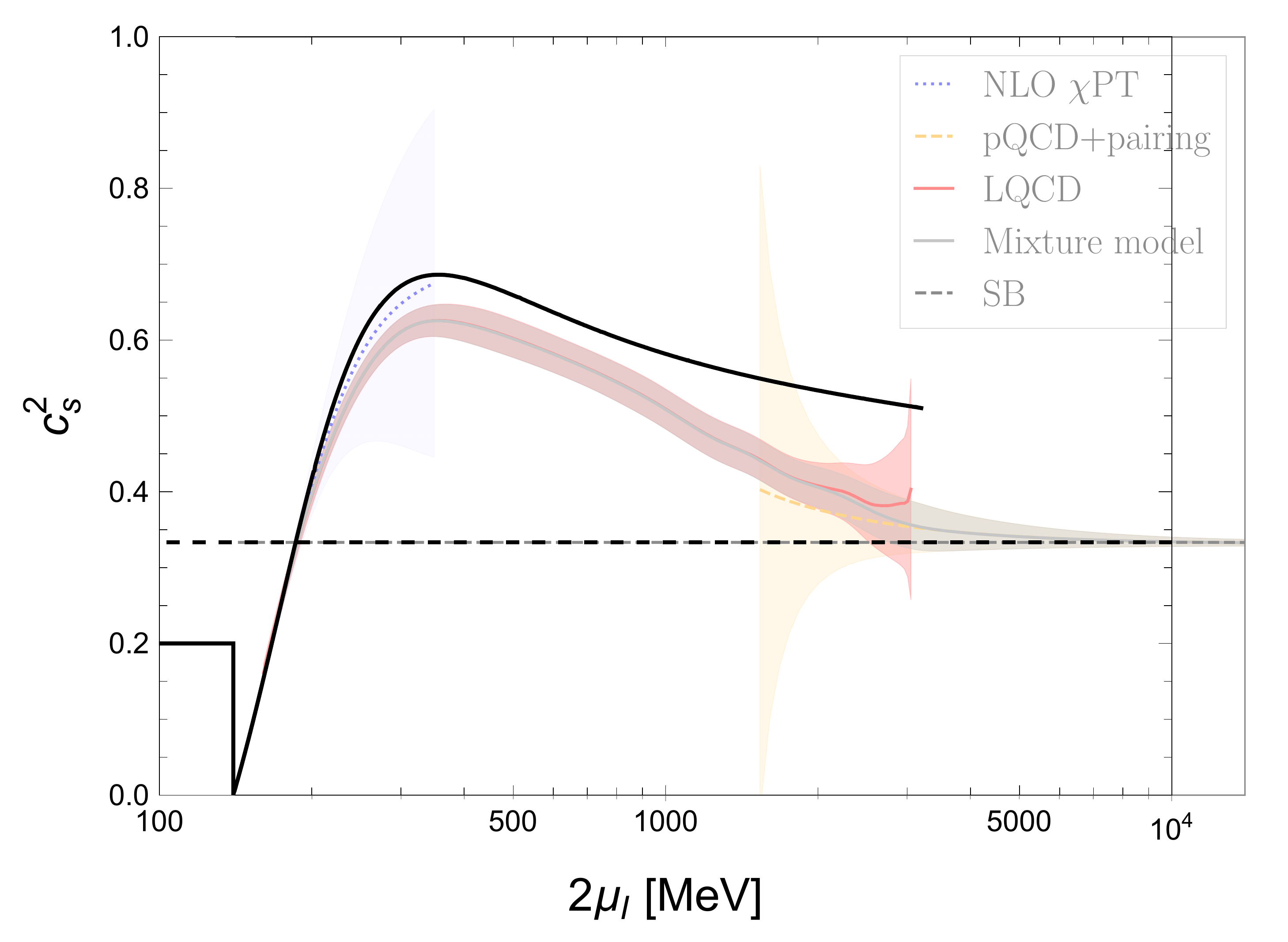}
\caption{Speed of sound squared at $T=0$ as a function of $\mu_I$, compared to the lattice results (red band) of ref.\ \cite{Abbott:2024vhj}. The holographic result asymptotes to $c_s^2=2/5$, in contrast to QCD, where $c_s^2\to 1/3$ (dashed line) due to asymptotic freedom. } 
\label{fig:sound0}
\end{center}
\end{figure}

In figure \ref{fig:sound0} we compare the speed of sound with the very recent lattice results of ref.\ \cite{Abbott:2024vhj}. Our result includes the isospin vacuum, where, in the limit $T\to 0$, we obtain $c_s^2=1/5$, which is independent of $m_\pi$ and thus the same as in the chiral limit  \cite{BitaghsirFadafan:2018uzs}. (One can check numerically that in the isospin vacuum the entropy goes like $T^5$ for small temperatures, with a prefactor that does depend on $m_\pi$.) This gives rise to  a discontinuity at the onset of pion condensation. Since the speed of sound involves second derivatives of the free energy with respect to $\mu_I$ and $T$, the discontinuity is a sign of a second-order transition. Like the result from the lattice, our curve exceeds the conformal value $c_s^2=1/3$ and has a maximum around 
$\mu_I\simeq 1.28 \, m_\pi$\footnote{Such a maximum above the conformal limit was also conjectured in QCD at nonzero {\it baryon} densities \cite{Tews:2018kmu}, where lattice results are unavailable. The WSS model supports this conjecture for baryonic \cite{BitaghsirFadafan:2018uzs,Kovensky:2021kzl} and quarkyonic \cite{Kovensky:2020xif} matter, albeit with a somewhat less pronounced maximum compared to the present result at nonzero isospin densities.}. Our speed of sound deviates from the lattice results already at smaller values of $\mu_I$ compared to the deviation of isospin density and energy density in figure \ref{fig:nIeps}. This observation emphasizes that the accuracy of our model strongly depends on the observable under consideration. While the speed of sound must approach the conformal limit 1/3 at asymptotically large $\mu_I$ in QCD, we find by extending the calculation of figure \ref{fig:sound0} that our result in the pion-condensed phase asymptotes to $c_s^2=2/5$. Interestingly, this is the same value found in the massless limit of the chirally symmetric phase \cite{BitaghsirFadafan:2018uzs}.

%%%%%%%%%%%%%%%%%%%%%%%%%%%%%%%%%%%%%%%%%%%%%%%%%%%%%%%%%
\subsection{Nonzero temperature}
\label{sec:results-Teffects}
%%%%%%%%%%%%%%%%%%%%%%%%%%%%%%%%%%%%%%%%%%%%%%%%%%%%%%%%%

Next, we evaluate our holographic model at nonzero temperatures. For all results in this section, the physical pion mass, i.e., the parameter set (\ref{Lfit}), is used because now the comparison with lattice QCD will be done with refs.\ \cite{Brandt:2017oyy,Brandt:2022hwy}, where the physical pion mass was used as well.

\begin{figure} [t]
\begin{center}
\includegraphics[width=0.50\textwidth]{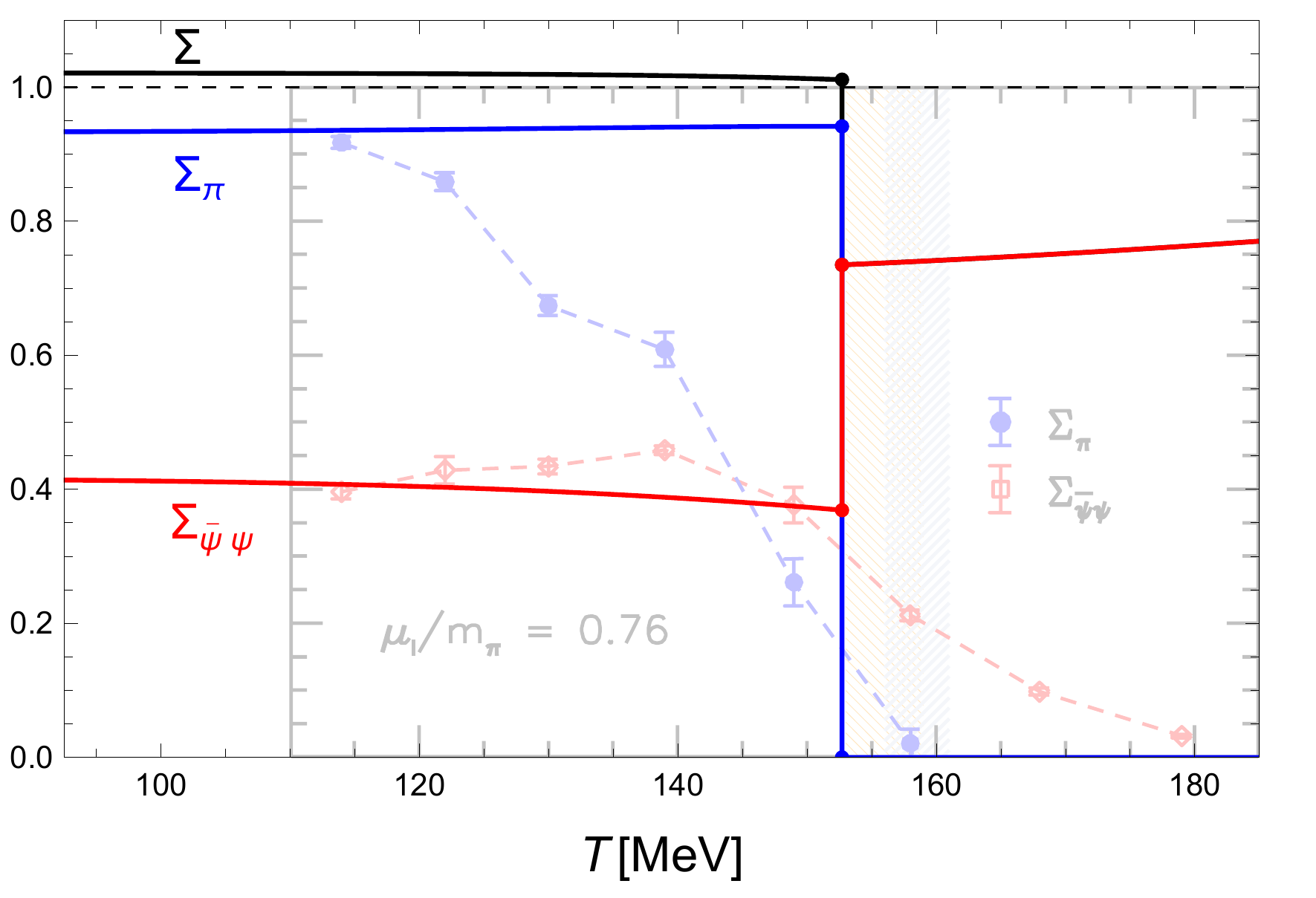}
\caption{Normalized pion and chiral condensates (\ref{SigSig}) as a function of $T$ at $\mu_I = 0.76\, m_\pi$, compared to the lattice results of ref.\ \cite{Brandt:2017oyy}. The black curve $\Sigma$ (\ref{constraint}) indicates the deviation from leading-order $\chi$PT, where this quantity is 1 (which we have indicated by the dashed line for comparison). After the chiral phase transition, where $\Sigma_\pi=0$, red and black lines coincide with each other.} 
\label{fig:SigSigT}
\end{center}
\end{figure}

We start with the condensates, i.e., the counterpart of figure \ref{fig:SigSigmu}. We fix $\mu_I=0.76 \, m_\pi$, as in the corresponding plot of ref.\ \cite{Brandt:2017oyy}, and show the  condensates in the vicinity of the chiral phase transition, see figure \ref{fig:SigSigT}. For the chosen isospin chemical potential, the pion-condensed phase is preferred at low temperatures. We see that both condensates $\Sigma_\pi$ and $\Sigma_{\bar{\psi}\psi}$ are in very good agreement with the lattice results at (and thus presumably for all temperatures up to) $T\simeq 115\, {\rm MeV}$. 

The behavior in the vicinity of the chiral phase transition, however, is very different. The lattice results suggest a continuous melting of the pion condensate until a second-order transition is reached. While the pion condensate starts melting, the chiral condensate first increases slightly. As already pointed out in ref.\ \cite{Brandt:2017oyy}, this might be a ``remnant'' of the constraint $\Sigma=1$ of leading-order $\chi$PT. Only after this increase, the chiral condensate decreases, showing the expected chiral crossover, at a temperature comparable to the critical temperature of pion condensation. In our holographic results it is, in contrast, the pion condensate that (very slightly) increases, while the chiral condensate decreases to maintain a roughly constant $\Sigma>1$. Then, there is a single critical temperature of a first-order transition at which both condensates are discontinuous. At this critical temperature, we find that the HTQ phase becomes the energetically preferred phase, taking over from the pion-condensed phase. We see that the critical temperature itself is in good agreement with the lattice, sitting at the lower end of the error bar for the pseudo-critical temperature of the chiral transition (see also the phase diagram in the next subsection). As expected, the pion condensate vanishes above the transition. 

The chiral condensate, however, assumes a {\it larger} value compared to temperatures below the critical one, although the ``total condensate'' $\Sigma$ decreases. As the temperature further increases, we even see a slight further increase of the chiral condensate, now identical to $\Sigma$. Most of this behavior can be traced back to known shortcomings of the model, unrelated to pion condensation. In particular, the unphysical order of the chiral phase transition in the WSS model (and in other top-down holographic models \cite{Evans:2011eu}) was observed in many studies before\footnote{A chiral crossover can be implemented -- by construction -- in bottom-up holographic models \cite{Rougemont:2023gfz}, including the V-QCD model \cite{Alho:2012mh}.}, and the issues with the definition of the chiral condensate based on the open Wilson line have been discussed for instance in ref.\ \cite{Kovensky:2019bih}, see also refs.\
\cite{Aharony:2008an,Hashimoto:2008sr,McNees:2008km,Argyres:2008sw}. 

\begin{figure} [t]
\begin{center}
\hbox{\includegraphics[width=0.48\textwidth]{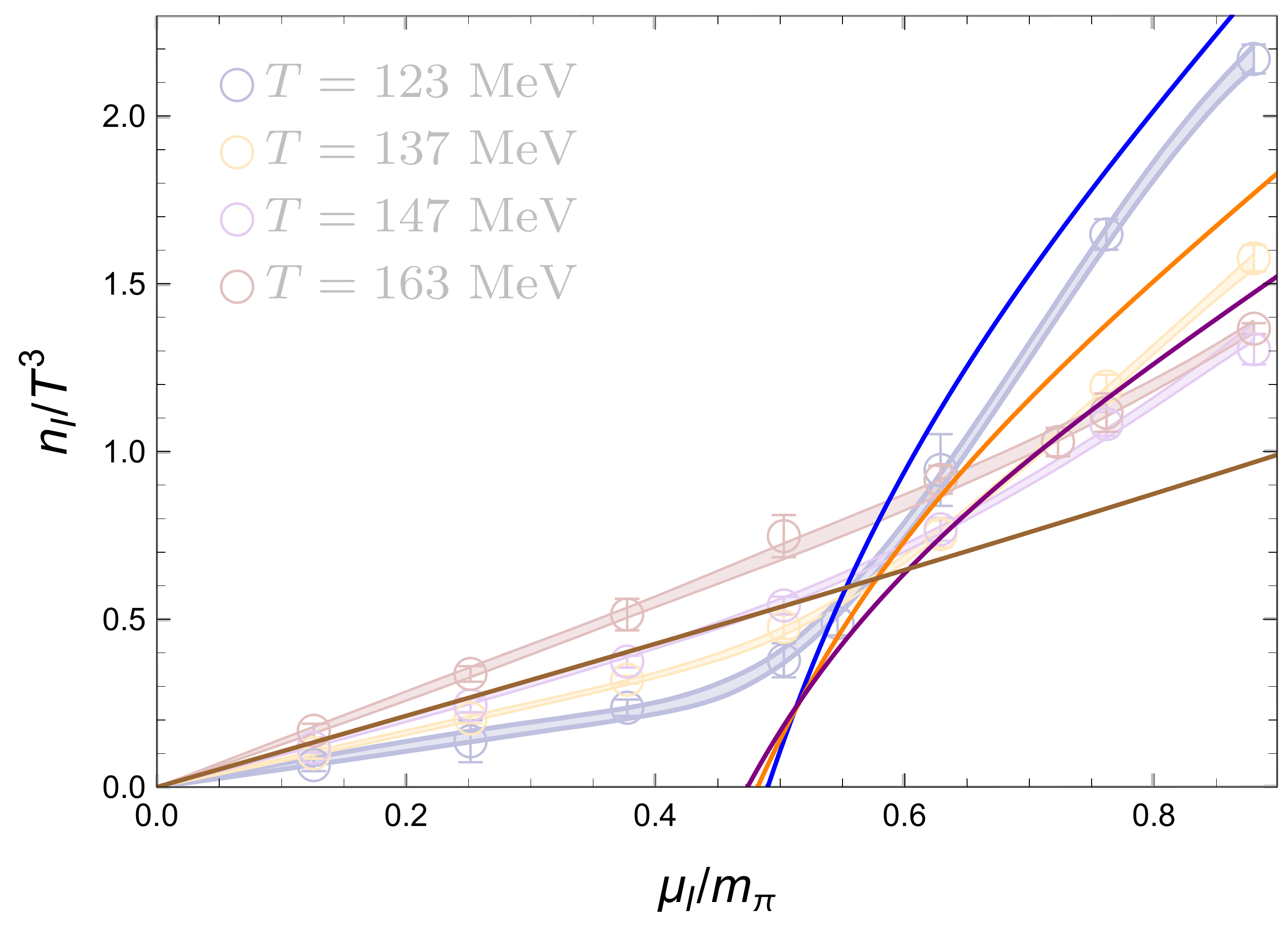}
\includegraphics[width=0.52\textwidth]{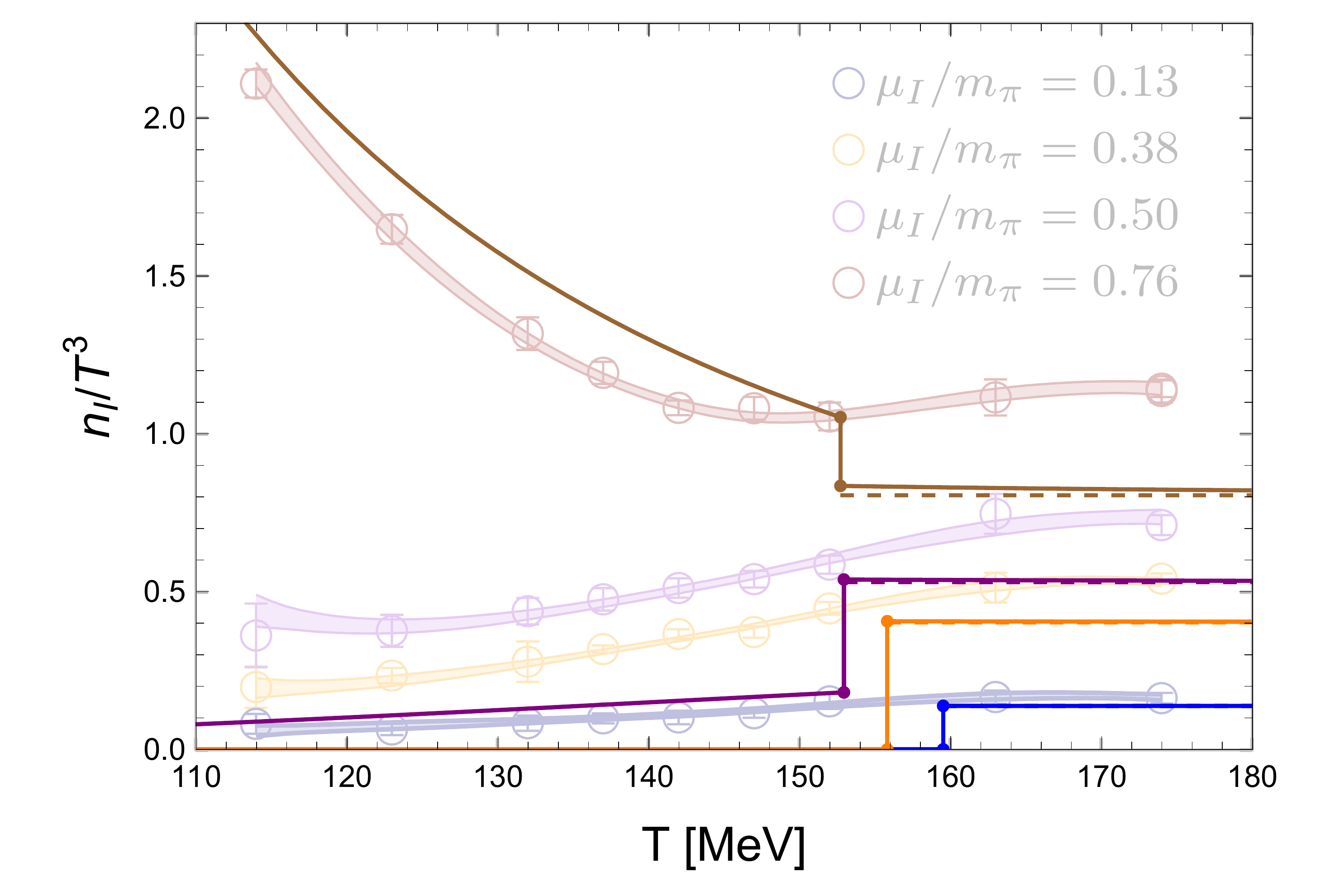}}
\caption{Isospin density (divided by $T^3$) as a function of  isospin chemical potential for four different temperatures (left), and as a function of temperature for four different isospin chemical potentials  (right). The results are compared with (and values of $T$ and $\mu_I$ chosen as in) figure 18 of ref.\ \cite{Brandt:2022hwy}. The dashed lines in the right panel correspond to the analytical result (\ref{nI massless HTQ large T}), valid in the chiral limit and small $\mu_I/T$.}  
\label{fig:nIT3}
\end{center}
\end{figure}

In figure \ref{fig:nIT3} we show the isospin density at fixed values of $T$ as a function of $\mu_I$ (left) and vice versa (right). As in ref.\ \cite{Brandt:2022hwy}, whose results are shown in the background for comparison, we show $n_I$ in units of $T^3$.
In the left panel, three of the temperatures, $T= (123,137,147) $ MeV are below the critical temperature of the chiral transition, such that, upon increasing $\mu_I$, the system transitions from the isospin vacuum to the pion-condensed phase. The fourth, $T= 163 $ MeV, is above the critical temperature, such that the system is in the HTQ phase for all $\mu_I$ in the shown range. 
In sharp contrast with lattice QCD simulations, our system remains -- even at nonzero $T$ -- in the isospin vacuum with $n_I$ being strictly zero until pion condensation sets in (or until the temperature is sufficiently large for the HTQ phase to be preferred). Using a weak-coupling formulation, our calculation does not account for a pion gas in this regime. This property is known from many previous holographic studies, mostly in the context of nonzero baryon chemical potential. 
It prevents the chiral transition from being a smooth crossover and is thus related to the above discussion of the order of the chiral transition. On the lattice, $n_I$ is already nonzero for small $\mu_I$, and the onset of pion condensation yields an additional increase of $n_I$ which is more and more cusp-like if the temperature is lowered.
In contrast, the holographic results show this cusp for any temperature below the critical one. After the cusp, in the presence of a pion condensate, the holographic results seem to overestimate the isospin density, while in the HTQ phase $n_I$ is smaller compared to the lattice calculation. We should  point out that Ref.~\cite{Brandt:2022hwy} includes strange quarks, which may account for parts of the quantitative discrepancy, at least for temperatures larger than the strange quark mass. 

In the right panel of figure \ref{fig:nIT3}, the two lowest chemical potentials, $\mu_I= (0.13, 0.38) \, m_\pi$, correspond to cases where there is no pion condensation at zero temperature. As in the left panel, we see that the isospin density in these cases is zero, this time up to the chiral phase transition. For the other two chemical potentials, $\mu_I= (0.50, 0.76) \, m_\pi$, the first part of the curves corresponds to the pion-condensed phase. Interestingly, in our model, $n_I/T^3$ jumps up at the phase transition as the temperature is increased for values of $\mu_I$ either below or  close to and above the pion onset, while it jumps down for larger isospin chemical potentials. At high temperatures, in the HTQ phase, our curves for $n_I/T^3$ are nearly flat, in qualitative, although not quantitative, agreement with the lattice. This behavior can be understood analytically. As explained in appendix \ref{App HTQ}, in the chiral limit and to leading order in $\mu_I/T$, the isospin density goes like $\mu_IT^3$. This is different from a naive dimensional analysis, which would suggest $\mu_IT^2$. In the holographic result, the ``missing'' dimension is provided by $L$, see eq.\ (\ref{nI massless HTQ large T}). We have included this analytical result (dashed lines) in comparison to the full numerical results. It turns out to be a very good approximation, even though  the numerical results are obtained for the physical pion mass and for sizable values of $\mu_I/T$ (for instance, $\mu_I/T\simeq 0.67$ for $\mu_I=0.76\, m_\pi$ and $T=160\, {\rm MeV}$). 

\begin{figure} [t]
\begin{center}
\includegraphics[width=0.50\textwidth]{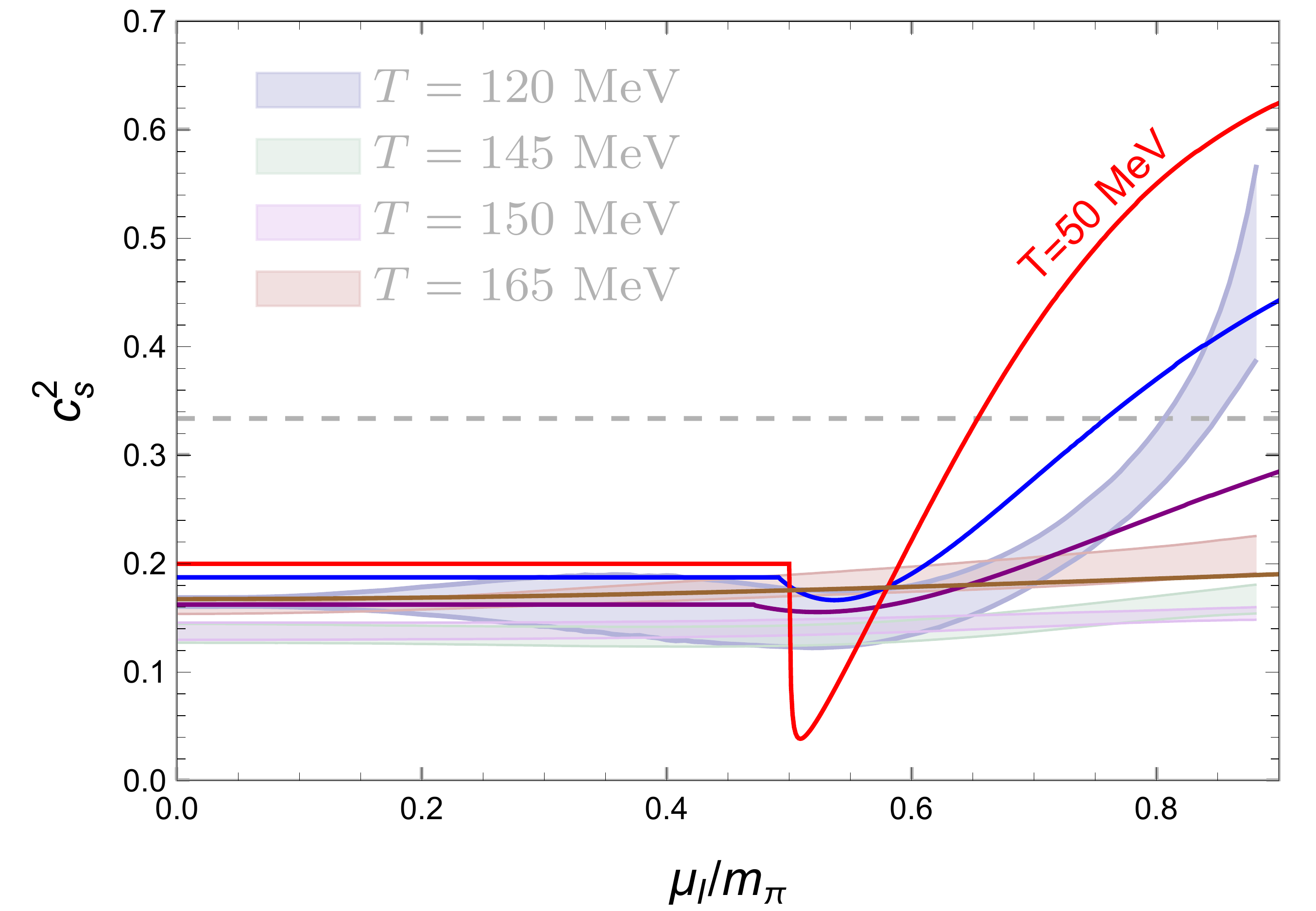}
\caption{Speed of sound squared as a function of isospin chemical potential at temperatures $T=(50,120,150,165)\, {\rm MeV}$ compared to the lattice results of ref.\ \cite{Brandt:2022hwy}. The same colors (blue, purple, brown) have been chosen for the three curves with identical temperatures compared to the lattice bands.} 
\label{fig:soundT}
\end{center}
\end{figure}

We show the speed of sound at nonzero temperatures in figure \ref{fig:soundT}, again compared to lattice results of ref.\ \cite{Brandt:2022hwy}. We have added a curve at $T=50$ MeV (red), not present in ref.\ \cite{Brandt:2022hwy}, to illustrate how the discontinuity at zero temperature, see figure \ref{fig:sound0}, turns into a cusp that becomes less and less pronounced as the temperature is increased. Conversely, to avoid too much cluttering, we have not included the $T=145$ MeV curve, as, in any case, it would have been very similar to the one for $T=150$ MeV. As in the left panel of figure \ref{fig:nIT3}, at the largest temperature shown here the system is in the HTQ phase, and thus the cusp from the onset of pion condensation is gone. We see that the conformal limit is exceeded also at nonzero -- not too large -- temperatures for sufficiently large $\mu_I$. The entire regime where $c_s^2>1/3$ is best illustrated with the help of the phase diagram, which we discuss now.

%%%%%%%%%%%%%%%%%%%%%%%%%%%%%%%%%%%%%%%%%%%%%%%%%%%%%%%%%%%%%%%%%%%%

%%%%%%%%%%%%%%%%%%%%%%%%%%%%%%%%%%%%%%%%%%%%%%%%%%%%%%%%%
\subsection{Phase diagram}
\label{sec:results-phasediagram}
%%%%%%%%%%%%%%%%%%%%%%%%%%%%%%%%%%%%%%%%%%%%%%%%%%%%%%%%%

\begin{figure} [t]
\begin{center}
\hbox{\includegraphics[width=0.512\textwidth]{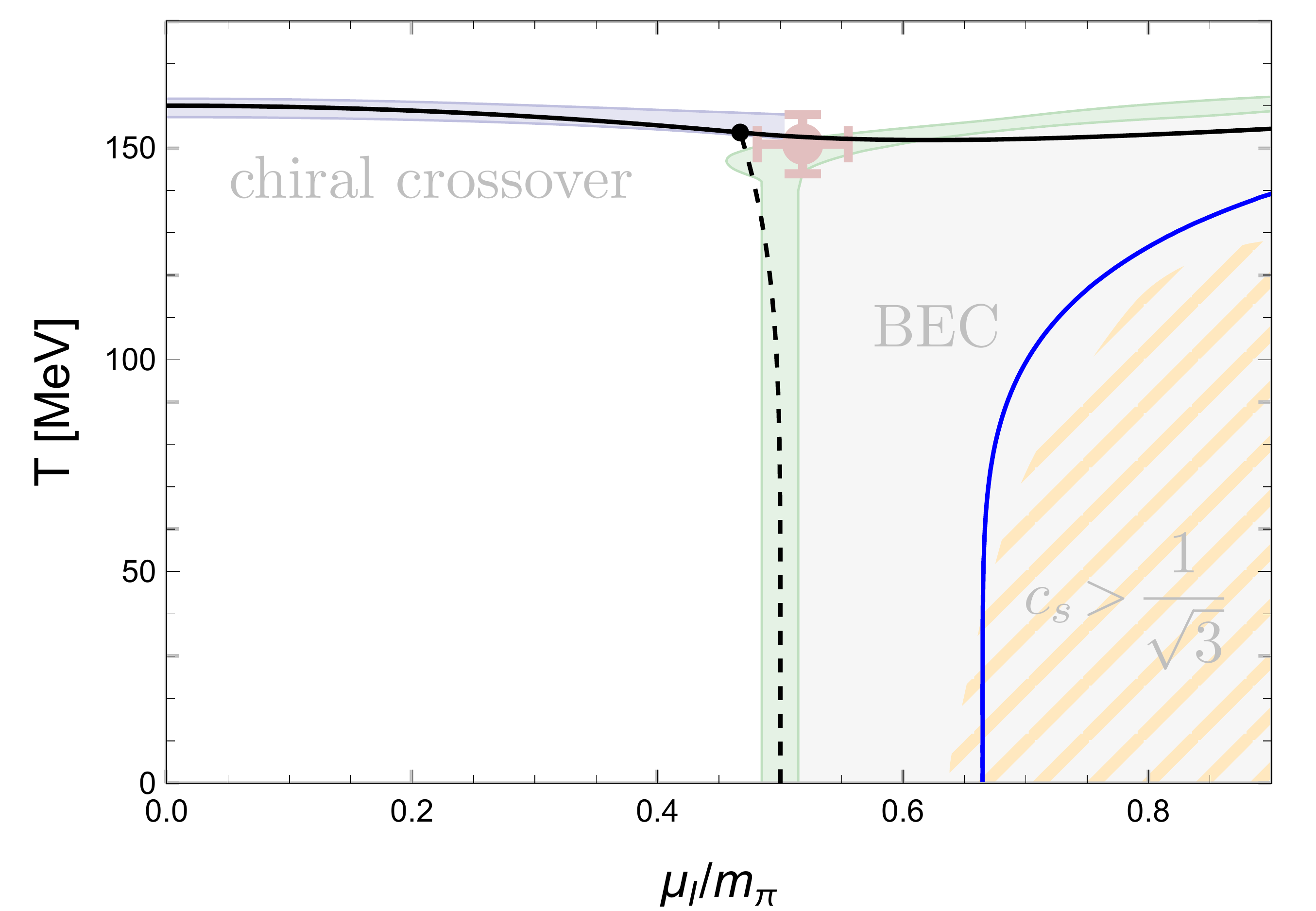}\includegraphics[width=0.488\textwidth]{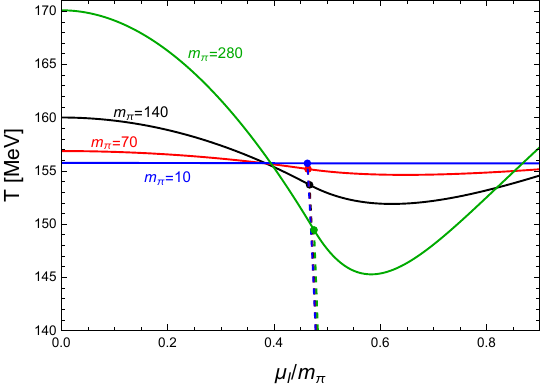}}
\caption{
{\it Left panel:} 
Phase diagram in the plane of isospin chemical potential and temperature. The holographic results (black and blue curves) are compared to lattice QCD results \cite{Brandt:2022hwy} (background). Black lines are first-order (solid) and second-order (dashed) phase transitions, the blue line is defined by $c_s^2=1/3$. The phase transition lines separate the HTQ phase (above the first-order line), the isospin vacuum (small $\mu_I$) and the pion-condensed phase (large $\mu_I$). 
{\it Right panel:}
Change of the phase transitions as the pion mass is varied.}
\label{fig:phasediagram}
\end{center}
\end{figure} 

The phase diagram in the $T$-$\mu_I$ plane is shown in 
the left panel of
figure \ref{fig:phasediagram}, in comparison with the phase diagram based on lattice QCD, taken from ref.\ \cite{Brandt:2022hwy}. Three out of our four possible phases populate the holographic phase diagram: the isospin vacuum, pion condensation, and the HTQ phase. The LTQ phase is always disfavored energetically. (If we were to ignore pion condensation, LTQ and HTQ phases would both account for the phases with nonzero $n_I$, analogous to the phase structure in the plane of temperature and baryon chemical potential of ref.\ \cite{Kovensky:2019bih}.)

The onset of pion condensation from the isospin vacuum is a second-order transition that ends in a tricritical point located at $(\mu_I,T)\simeq (0.47 \, m_\pi,  154 \, {\rm MeV})$. Its $T$ dependence is remarkably mild, but for large temperatures we see that the critical chemical potential decreases slightly, thereby increasing the region of pion condensation. This is somewhat surprising since temperature effects typically act against condensation (which, eventually, they also do here). It is less surprising, however, if we recall figure \ref{fig:SigSigT}, where we have already observed that the pion condensate can slightly increase with temperature, counteracting the tendency of the chiral condensate. 

The first-order transition where the pion onset curve terminates separates the HTQ phase from the isospin vacuum and from the pion-condensed phase and is thus interpreted as the chiral transition. (This is most obvious in the chiral limit, where the HTQ phase contains straight, disconnected flavor branes and describes chirally symmetric, massless quarks.) The details of this 
transition were already discussed with the help of figures \ref{fig:SigSigT} and \ref{fig:nIT3}. Here, in the phase diagram, we see that the variation of this curve with $\mu_I$ is very mild, at least in the range shown here, showing a shallow minimum just after the onset of pion condensation. Extending the range to (much) larger values of $\mu_I$, we find that the transition temperature increases until it appears to saturate at about $T_c\sim 300\, {\rm MeV}$. This is in accordance with earlier results in the WSS model \cite{Kovensky:2021ddl}, where the chiral limit was considered, which is a good approximation for $\mu_I\gg m_\pi$. In any case, as already discussed above, we do not trust the model to give much insight into QCD at these ultra-high isospin densities. 

Our phase diagram also includes an additional curve (blue), which is not a phase transition but is defined by $c_s^2=1/3$. To the right of the curve, i.e., for larger chemical potentials, the speed of sound exceeds the conformal limit, see also figures \ref{fig:sound0} and \ref{fig:soundT}. This curve, while also almost vertical for small $T$, does show a stronger temperature sensitivity compared to the onset of pion condensation.  

Let us comment on the comparison with the lattice results. We have already mentioned the difference compared to lattice QCD in the nature of our phase transitions and in some of our observables at  large temperatures. The overall phase structure, however, including the region of the super-conformal speed of sound, is in remarkably good agreement with lattice QCD. We recall that our 3 model parameters are chosen to reproduce, besides the pion decay constant, the two points where the phase transition lines meet the axes. Thus, the nontrivial agreement with lattice QCD is mainly manifest in the T-shaped form of the phase structure and the resulting location of our tricritial point. In particular, the nearly vertical pion onset is difficult to reproduce in $\chi$PT \cite{Adhikari:2020kdn} and previous holographic studies \cite{Cao:2020ske}. It has also been observed in a Polyakov-loop quark-meson model \cite{Adhikari:2018cea,Chiba:2024cny}, which, however, has many more knobs to turn in the form of model parameters (in that model, the pion onset curve tends to bend slightly to larger $\mu_I$ at large $T$, in contrast to our results). 

In the right panel of figure \ref{fig:phasediagram} we show our phase transition lines for half and twice the physical pion mass and a very small pion mass of $10\, {\rm MeV}$. This plot shows that while the phase transition at $\mu_I=0$ decreases with $m_\pi$, the temperature of the tricritical point {\it increases}, rendering the phase transition lines more and more straight as we approach the chiral limit. It would be interesting to see if this prediction is borne out in future lattice calculations.

%%%%%%%%%%%%%%%%%%%%%%%%%%%%%%%%%%%%%%%%%%%%%%%%%%%%%%%%%
\section{Summary and outlook}
\label{sec:outlook}

%%%%%%%%%%%%%%%%%%%%%%%%%%%%%%%%%%%%%%%%%%%%%%%%%%%%%%%%%

We have studied pion condensation at nonzero temperature and isospin chemical potential in the holographic WSS model. More precisely, we have worked in the decompactified limit  of the model, where the geometry of the flavor branes is dynamically computed in the presence of the thermal medium and the pion condensate. This calculation is done for the first time in the presence of an effective mass term in the action, which has allowed us to work with the physical pion mass. 

We have presented a detailed comparison of our results for chiral and pion condensates, thermodynamic quantities, and the speed of sound with $\chi$PT and, more importantly, with recent results from lattice QCD. We have shown that our model reproduces leading-order $\chi$PT for small isospin densities and temperatures. In comparison to lattice QCD, we have pointed out very good agreement of the low-temperature thermodynamics, including the speed of sound, and also of the overall phase structure in the $T$-$\mu_I$ plane. This is remarkable in view of the small number of model parameters: the 't Hooft coupling, the pion mass, and the length scale for the asymptotic separation of the flavor branes. 

We also have pointed out discrepancies to the lattice results, which are difficult to avoid in an approximation that relies on infinite $N_c$ and $\lambda$ before the results are extrapolated to $N_c=3$ and finite coupling strength. These discrepancies concern for instance the order of the chiral phase transition and the chiral condensate itself, as well as the behavior of certain thermodynamic quantities at large temperatures. These discrepancies serve as a guidance for future studies regarding the improvement of the model. For instance, the renormalization of the Nambu-Goto action in the effective mass term could be revisited to obtain a more realistic definition of the chiral condensate. Another interesting question is whether the model can be improved to allow for a chiral crossover instead of a first-order phase transition without including full-fledged string corrections or a completely dynamical geometry, which is difficult.    

The WSS model was already employed in numerous studies in the context of QCD thermodynamics and phase structure, including applications to neutron stars. Our work has been the first to perform a systematic comparison with lattice calculations, and the encouraging agreement with the phase diagram can be viewed as an {\it a posteriori} validation of these earlier works, keeping in mind the shortcomings just mentioned. One should also keep in mind that applications to dense matter in neutron stars requires an  approximation for nuclear matter, and a realistic $T$-$\mu_I$ phase diagram does of course not guarantee that these baryonic approximations are realistic as well. 

One natural question raised by our study is whether we can use our setup to make predictions for isospin QCD itself that have not been settled or even been addressed by lattice calculations (besides applying the model to experimentally accessible regimes that usually involve baryon density). An interesting problem is the conjectured existence of a BEC-BCS crossover in the $T$-$\mu_I$ plane. While we do not expect our model to go all the way to the weakly coupled BCS regime, we have pointed out a conceivable configuration that allows for the coexistence of the pion condensate with quark degrees of freedom. It would be interesting to investigate whether this configuration is indeed relevant for the phase structure and whether it gives us a prediction for the location of the BEC-BCS crossover region \cite{progress}.

%%%%%%%%%%%%%%%%%%%%%%%%%%%%%%%%%%%%%%%%%%%%%%%%%%%%%%%%%
\acknowledgments

We thank Ryan Abbott, Mark Alford, Lorenzo Bartolini, Bastian Brandt, Gergely Endr\H{o}di, and Savvas Pitsinigkos
for useful comments and discussions. 

\appendix

\section{Phases without pion condensation}
\label{AppA}

In this appendix, we briefly discuss the phases without pion condensation, $\theta=0$, and collect the relevant equations needed for our main results. All these phases were either discussed already in the previous literature or are isospin analogues of phases discussed previously with nonzero baryon chemical potential. Without pion condensation, we have for all phases in this appendix $K_1(u)=K_2(u)=0$, and the boundary condition $K_3(\infty) = \bar{\mu}_I$.

%%%%%%%%%%%%%%%%%%%%%%%%%%%%%%%%%%%%%
\subsection{Isospin vacuum}
\label{App IVac}
%%%%%%%%%%%%%%%%%%%%%%%%%%%%%%%%%%%%%

This phase is identical to the mesonic phase of refs.\ \cite{Kovensky:2019bih,Kovensky:2020xif}. One can obtain this phase by taking the limit $\theta=\bar{n}_I=0$ in the pion-condensed phase of section \ref{sec:pi-phase}. More explicitly, the gauge potential is constant, $K_{3}(u) = \bar{\mu}_I$, and the embedding function is given by
\bea 
x_4' = \frac{k+A\phi_T}{u^{11/2}f_T}\zeta \, , 
\eea
with
\be
\zeta = \sqrt{1+u^3f_Tx_4'^2} = \left[1-\frac{(k+A\phi_T)^2}{u^8f_T}\right]^{-1/2} \, ,
\ee
and
\be
\label{k value vacuum}
k = u_c^4\sqrt{f_T(u_c)}-A\phi_T(u_c) \, .
\ee
For the numerical evaluation, we need to solve the coupled equations coming from the boundary condition of $x_4$ and the self-consistency condition for $A$,  
\begin{subequations}
\label{ell2Mes}
\bea
\frac{\ell}{2} &=& \int_{u_c}^\infty du\, x_4' \, , \\[2ex]
A &=& \frac{2\alpha}{\lambda_0^2}\exp\left(2\lambda_0\int_{u_c}^\infty du\, \phi_T x_4'\right) \, , \label{AIvac}
\eea
\end{subequations}
for the variables $A$ and $u_c$. String sources are absent in this phase, hence the free energy is independent of $\mu_I$ and can be written as 
\be
\label{OM IVac}
\Omega = \int_{u_c}^\infty du\, u^{5/2}\zeta - \frac{A}{2\lambda_0}  \,.
\ee
The speed of sound is computed as explained in section \ref{sec:sound}, where, in the isospin vacuum, the entropy density assumes the form
\bea
s &=& \frac{2}{t}\int_{u_c}^\infty du\, x_4'\left[\frac{3u_T^3}{2u^3}\frac{k+A\phi_T}{f_T}+A\left(\phi_T-\frac{u}{\sqrt{f_T}}\right)\right]  \, .
\eea

%%%%%%%%%%%%%%%%%%%%%%
\subsection{Low-temperature quark phase (LTQ)}
\label{App LTQ}
%%%%%%%%%%%%%%%%%%%%%%

In this phase, the flavor branes also join at the point denoted by $u_c$. At this point, strings are attached and a cusp develops in the embedding. The (integrated) equations of motion can be written as  
\be
K_3'=\frac{\bar{n}_I}{u^{5/2}}\zeta \, , \qquad x_4'=\frac{A\phi_T+k}{u^{11/2}f_T}\zeta \, , \label{eomLTQ}
\ee
with 
\be
\zeta = \sqrt{1+u^3f_Tx_4'^2-K_3'^2} = \left[1-\frac{(k+A\phi_T)^2}{u^8f_T}+\frac{\bar{n}_I^2}{u^5}\right]^{-1/2} \, . \label{zetaLTQ}
\ee
The free energy becomes
\begin{align}
\Omega &= \int_{u_c}^\infty du\, u^{5/2}\zeta+\bar{n}_I[u_c-u_T-K_3(u_c)]-\frac{A}{2\lambda_0}  \nn \\[2ex]
&=  \int_{u_c}^\infty du\left(\frac{u^{5/2}}{\zeta}+A\phi_Tx_4'\right)-\bar{\mu}_I\bar{n}_I+k\frac{\ell}{2}+\bar{n}_I(u_c-u_T)-\frac{A}{2\lambda_0} \, . \label{omegaLTQ}
\end{align}
We find $k = u_c^4\sqrt{f_T(u_c)}-A\phi_T(u_c)$ from the minimization with respect to $u_c$, the same result as in the isospin vacuum. We also obtain 
\be
0 = \frac{\partial\Omega}{\partial \bar{n}_I} = u_c-u_T-K_3(u_c)  \, .
\ee
Hence, for given $\mu_I$ and $T$ we need to solve the coupled equations 
\be 
\frac{\ell}{2} = \int_{u_c}^\infty du\, x_4' \,, 
\qquad A = \frac{2\alpha}{\lambda_0^2}\exp\left(2\lambda_0\int_{u_c}^\infty du\, \phi_T x_4'\right) \, , \qquad 
\bar{\mu}_I =  \int_{u_c}^\infty du\, K_3'+u_c-u_T 
\, ,
\ee
for the variables $A$, $u_c$, $\bar{n}_I$ and insert the results back into eq.\ (\ref{omegaLTQ}) to obtain the free energy.  
Again,  one can write down a semi-analytical expression for the entropy density, in order to calculate the speed of sound,
\begin{equation}
    s = \frac{2}{t} \left\{
    \bar{n}_I u_T + \int_{u_c}^{\infty} du \, x_4' \left[
    \frac{3 u_T^3}{2 u^3} \frac{A \phi_T + k}{f_T} + A \left(
    \phi_T - \frac{u}{\sqrt{f_T}}
    \right)
    \right]
    \right\} \, .
\end{equation}

%%%%%%%%%%%%%%%%%%%%%
\subsection{High-temperature quark phase (HTQ)}
\label{App HTQ}
%%%%%%%%%%%%%%%%%%%%%

If the strings of the LTQ phase are hidden behind the horizon, they pull the flavor branes all the way down to $u_T$, hence separating them (they become straight in the chiral limit). This leads to the HTQ phase, with the boundary condition $K_3(u_T)=0$, and where the integrated equations of motion can be written in exactly the same form as for the LTQ phase, i.e., we may use eqs.\ (\ref{eomLTQ}) and (\ref{zetaLTQ}). However, the expression of the Nambu-Goto action in $A$ is different due to the different shape of the embedding, 
\be \label{AHTQ}
A = \frac{2\alpha}{\lambda_0^2}\exp\left\{2\lambda_0\left[\phi_T(u_T)x_4(u_T)+\int_{u_T}^\infty du\, \phi_T x_4'\right]\right\} \, ,
\ee
where 
\be
\phi_T(u_T)=\frac{3\sqrt{\pi}\,\Gamma[5/3]}{2\Gamma[1/6]}\, u_T \, .
\ee
The free energy can be written as 
\begin{align}
\Omega & = \int_{u_T}^\infty du\, u^{5/2}\zeta -\frac{A}{2\lambda_0} \nn \\[2ex]
&= \int_{u_T}^\infty du\left(\frac{u^{5/2}}{\zeta}+A\phi_Tx_4'\right)-\frac{A}{2\lambda_0}-\bar{\mu}_I\bar{n}_I+k\left[\frac{\ell}{2}-x_4(u_T)\right] \, .
\label{OM HTQ}
\end{align}
Stationarity of $\Omega$ with respect to $k$ and $x_4(u_T)$ give, respectively, the boundary condition 
\be \label{ell2HTQ}
\frac{\ell}{2} = \int_{u_T}^\infty du\, x_4' +x_4(u_T) \, ,
\ee
and the value of $k$,
\be
k = -A\phi_T(u_T) \, .
\ee
For the numerical evaluation, we need to solve eq.\ (\ref{AHTQ}) together with 
\be
\bar{\mu}_I = \int_{u_T}^\infty du\, K_3' \, , 
\ee
for $\bar{n}_I$ and $A$. One can then compute $x_4(u_T)$ from \eqref{ell2HTQ} if needed. Of course, only solutions where $x_4(u_T) \geq 0$ are physical. In particular, for $x_4(u_T) = 0$, the HTQ configuration connects to that of the LTQ phase. The entropy density, needed for the speed of sound, becomes 
\begin{eqnarray}
    s &=& \frac{2}{t} \left\{ A \phi_T(u_T) x_4(u_T) +
    u_T^{7/2} \sqrt{1+\frac{\bar{n}_I^2}{u_T^5} - \frac{4A^2}{9u_T^6}}
    \right. \nn \\[2ex]
    &&\left. + \int_{u_T}^{\infty} du \, x_4' \left[
    \frac{3 u_T^3}{2 u^3} \frac{A \phi_T + k}{f_T} + A \left(
    \phi_T - \frac{u}{\sqrt{f_T}}
    \right)
    \right]
    \right\} \, .
\end{eqnarray}

In the massless limit, the HTQ phase simplifies drastically \cite{Horigome:2006xu,Kovensky:2019bih,Kovensky:2021ddl} because the embedding function becomes constant, $x_4(u) = \ell/2$. In that case, the relation between isospin density, isospin chemical potential, and temperature is  \cite{Kovensky:2021ddl} 
\begin{equation}
\label{muI massless HTQ}
    \bar{\mu}_I = \frac{\Gamma[3/10]\Gamma[6/5]}{\sqrt{\pi}}\,\bar{n}_I^{2/5} - u_T \, 
    {}_2F_1\left[\frac{1}{5},\frac{1}{2},\frac{6}{5},-\frac{u_T^5}{\bar{n}_I^2}\right] \, .
\end{equation}
In general, this equation needs to be solved numerically. Here we are interested in an analytical result for small chemical potentials. To this end, we may solve eq.\ (\ref{muI massless HTQ}) for $\bar{n}_I^{2/5}/u_T$ in a series expansion in 
$\bar{\mu}_I/u_T$. After reinstating all constants to obtain dimensionful quantities, the leading term for the isospin density is 
\begin{equation}
\label{nI massless HTQ large T}
    n_I = \frac{32\pi \tilde{\lambda}_0 L}{9} \mu_I T^3 + \ldots \, .  
\end{equation}
This expression turns out to be a good approximation for the results discussed in section \ref{sec:results-Teffects}.

\bibliographystyle{JHEP}
\bibliography{references}

\providecommand{\href}[2]{#2}\begingroup\raggedright\begin{thebibliography}{10}

\bibitem{GasserLeutwyler:1983CPT}
J.~Gasser and H.~Leutwyler, {\it {Chiral Perturbation Theory to One Loop}},  {\em Annals of Phys.} {\bf 158} (1984) 142--210.

\bibitem{Ecker:1994gg}
G.~Ecker, {\it {Chiral perturbation theory}},  {\em Prog. Part. Nucl. Phys.} {\bf 35} (1995) 1--80, [\href{http://arxiv.org/abs/hep-ph/9501357}{{\tt hep-ph/9501357}}].

\bibitem{Gell-Mann:1960mvl}
M.~Gell-Mann and M.~Levy, {\it {The axial vector current in beta decay}},  {\em Nuovo Cim.} {\bf 16} (1960) 705.

\bibitem{Lee:1968da}
B.~W. Lee, {\it {Renormalization of the sigma model}},  {\em Nucl. Phys. B} {\bf 9} (1969) 649--672.

\bibitem{Nambu:1961tp}
Y.~Nambu and G.~Jona-Lasinio, {\it {Dynamical Model of Elementary Particles Based on an Analogy with Superconductivity. 1.}},  {\em Phys. Rev.} {\bf 122} (1961) 345--358.

\bibitem{Nambu:1961fr}
Y.~Nambu and G.~Jona-Lasinio, {\it {Dynamical model of elementary particles based on an analogy with superconductivity. II}},  {\em Phys.Rev.} {\bf 124} (1961) 246--254.

\bibitem{Maldacena:1997re}
J.~M. Maldacena, {\it {The Large N limit of superconformal field theories and supergravity}},  {\em Adv.Theor.Math.Phys.} {\bf 2} (1998) 231--252, [\href{http://arxiv.org/abs/hep-th/9711200}{{\tt hep-th/9711200}}].

\bibitem{Gubser:1998bc}
S.~S. Gubser, I.~R. Klebanov, and A.~M. Polyakov, {\it {Gauge theory correlators from noncritical string theory}},  {\em Phys. Lett. B} {\bf 428} (1998) 105--114, [\href{http://arxiv.org/abs/hep-th/9802109}{{\tt hep-th/9802109}}].

\bibitem{Witten:1998qj}
E.~Witten, {\it {Anti-de Sitter space and holography}},  {\em Adv. Theor. Math. Phys.} {\bf 2} (1998) 253--291, [\href{http://arxiv.org/abs/hep-th/9802150}{{\tt hep-th/9802150}}].

\bibitem{Witten:1998zw}
E.~Witten, {\it {Anti-de Sitter space, thermal phase transition, and confinement in gauge theories}},  {\em Adv.Theor.Math.Phys.} {\bf 2} (1998) 505--532, [\href{http://arxiv.org/abs/hep-th/9803131}{{\tt hep-th/9803131}}].

\bibitem{Sakai:2004cn}
T.~Sakai and S.~Sugimoto, {\it {Low energy hadron physics in holographic QCD}},  {\em Prog. Theor. Phys.} {\bf 113} (2005) 843--882, [\href{http://arxiv.org/abs/hep-th/0412141}{{\tt hep-th/0412141}}].

\bibitem{Sakai:2005yt}
T.~Sakai and S.~Sugimoto, {\it {More on a holographic dual of QCD}},  {\em Prog. Theor. Phys.} {\bf 114} (2005) 1083--1118, [\href{http://arxiv.org/abs/hep-th/0507073}{{\tt hep-th/0507073}}].

\bibitem{Son:2000xc}
D.~T. Son and M.~A. Stephanov, {\it {QCD at finite isospin density}},  {\em Phys. Rev. Lett.} {\bf 86} (2001) 592--595, [\href{http://arxiv.org/abs/hep-ph/0005225}{{\tt hep-ph/0005225}}].

\bibitem{Son:2000by}
D.~T. Son and M.~A. Stephanov, {\it {QCD at finite isospin density: From pion to quark - anti-quark condensation}},  {\em Phys. Atom. Nucl.} {\bf 64} (2001) 834--842, [\href{http://arxiv.org/abs/hep-ph/0011365}{{\tt hep-ph/0011365}}].

\bibitem{Splittorff:2000mm}
K.~Splittorff, D.~T. Son, and M.~A. Stephanov, {\it {QCD - like theories at finite baryon and isospin density}},  {\em Phys. Rev. D} {\bf 64} (2001) 016003, [\href{http://arxiv.org/abs/hep-ph/0012274}{{\tt hep-ph/0012274}}].

\bibitem{Kogut:2002tm}
J.~B. Kogut and D.~K. Sinclair, {\it {Quenched lattice QCD at finite isospin density and related theories}},  {\em Phys. Rev. D} {\bf 66} (2002) 014508, [\href{http://arxiv.org/abs/hep-lat/0201017}{{\tt hep-lat/0201017}}].

\bibitem{Kogut:2002zg}
J.~B. Kogut and D.~K. Sinclair, {\it {Lattice QCD at finite isospin density at zero and finite temperature}},  {\em Phys. Rev. D} {\bf 66} (2002) 034505, [\href{http://arxiv.org/abs/hep-lat/0202028}{{\tt hep-lat/0202028}}].

\bibitem{Brandt:2017oyy}
B.~B. Brandt, G.~Endr\H{o}di, and S.~Schmalzbauer, {\it {QCD phase diagram for nonzero isospin-asymmetry}},  {\em Phys. Rev. D} {\bf 97} (2018), no.~5 054514, [\href{http://arxiv.org/abs/1712.08190}{{\tt arXiv:1712.08190}}].

\bibitem{Brandt:2018omg}
B.~B. Brandt and G.~Endr\H{o}di, {\it {Reliability of Taylor expansions in QCD}},  {\em Phys. Rev. D} {\bf 99} (2019), no.~1 014518, [\href{http://arxiv.org/abs/1810.11045}{{\tt arXiv:1810.11045}}].

\bibitem{Cuteri:2021hiq}
F.~Cuteri, B.~B. Brandt, and G.~Endr\H{o}di, {\it {Searching for the BCS phase at nonzero isospin asymmetry}},  {\em PoS} {\bf LATTICE2021} (2022) 232, [\href{http://arxiv.org/abs/2112.11113}{{\tt arXiv:2112.11113}}].

\bibitem{Brandt:2022hwy}
B.~B. Brandt, F.~Cuteri, and G.~Endr\H{o}di, {\it {Equation of state and speed of sound of isospin-asymmetric QCD on the lattice}},  {\em JHEP} {\bf 07} (2023) 055, [\href{http://arxiv.org/abs/2212.14016}{{\tt arXiv:2212.14016}}].

\bibitem{Abbott:2023coj}
{\bf NPLQCD} Collaboration, R.~Abbott, W.~Detmold, F.~Romero-L\'opez, Z.~Davoudi, M.~Illa, A.~Parre\~no, R.~J. Perry, P.~E. Shanahan, and M.~L. Wagman, {\it {Lattice quantum chromodynamics at large isospin density}},  {\em Phys. Rev. D} {\bf 108} (2023), no.~11 114506, [\href{http://arxiv.org/abs/2307.15014}{{\tt arXiv:2307.15014}}].

\bibitem{Abbott:2024vhj}
R.~Abbott, W.~Detmold, M.~Illa, A.~Parre\~no, R.~J. Perry, F.~Romero-L\'opez, P.~E. Shanahan, and M.~L. Wagman, {\it {QCD constraints on isospin-dense matter and the nuclear equation of state}},  \href{http://arxiv.org/abs/2406.09273}{{\tt arXiv:2406.09273}}.

\bibitem{Kojo:2024sca}
T.~Kojo, D.~Suenaga, and R.~Chiba, {\it {Isospin QCD as a laboratory for dense QCD}},  \href{http://arxiv.org/abs/2406.11059}{{\tt arXiv:2406.11059}}.

\bibitem{Hidaka:2011jj}
Y.~Hidaka and N.~Yamamoto, {\it {No-Go Theorem for Critical Phenomena in Large-Nc QCD}},  {\em Phys. Rev. Lett.} {\bf 108} (2012) 121601, [\href{http://arxiv.org/abs/1110.3044}{{\tt arXiv:1110.3044}}].

\bibitem{Fujimoto:2023unl}
Y.~Fujimoto and S.~Reddy, {\it {Bounds on the equation of state from QCD inequalities and lattice QCD}},  {\em Phys. Rev. D} {\bf 109} (2024), no.~1 014020, [\href{http://arxiv.org/abs/2310.09427}{{\tt arXiv:2310.09427}}].

\bibitem{Cohen:2003ut}
T.~D. Cohen, {\it {QCD inequalities for the nucleon mass and the free energy of baryonic matter}},  {\em Phys. Rev. Lett.} {\bf 91} (2003) 032002, [\href{http://arxiv.org/abs/hep-ph/0304024}{{\tt hep-ph/0304024}}].

\bibitem{Bergman:2007wp}
O.~Bergman, G.~Lifschytz, and M.~Lippert, {\it {Holographic Nuclear Physics}},  {\em JHEP} {\bf 11} (2007) 056, [\href{http://arxiv.org/abs/0708.0326}{{\tt arXiv:0708.0326}}].

\bibitem{Parnachev:2007bc}
A.~Parnachev, {\it {Holographic QCD with Isospin Chemical Potential}},  {\em JHEP} {\bf 02} (2008) 062, [\href{http://arxiv.org/abs/0708.3170}{{\tt arXiv:0708.3170}}].

\bibitem{Aharony:2007uu}
O.~Aharony, K.~Peeters, J.~Sonnenschein, and M.~Zamaklar, {\it {Rho meson condensation at finite isospin chemical potential in a holographic model for QCD}},  {\em JHEP} {\bf 02} (2008) 071, [\href{http://arxiv.org/abs/0709.3948}{{\tt arXiv:0709.3948}}].

\bibitem{Rebhan:2008ur}
A.~Rebhan, A.~Schmitt, and S.~A. Stricker, {\it {Meson supercurrents and the Meissner effect in the Sakai-Sugimoto model}},  {\em JHEP} {\bf 05} (2009) 084, [\href{http://arxiv.org/abs/0811.3533}{{\tt arXiv:0811.3533}}].

\bibitem{Preis:2010cq}
F.~Preis, A.~Rebhan, and A.~Schmitt, {\it {Inverse magnetic catalysis in dense holographic matter}},  {\em JHEP} {\bf 1103} (2011) 033, [\href{http://arxiv.org/abs/1012.4785}{{\tt arXiv:1012.4785}}].

\bibitem{Preis:2011sp}
F.~Preis, A.~Rebhan, and A.~Schmitt, {\it {Holographic baryonic matter in a background magnetic field}},  {\em J. Phys. G} {\bf 39} (2012) 054006, [\href{http://arxiv.org/abs/1109.6904}{{\tt arXiv:1109.6904}}].

\bibitem{Preis:2012fh}
F.~Preis, A.~Rebhan, and A.~Schmitt, {\it {Inverse magnetic catalysis in field theory and gauge-gravity duality}},  {\em Lect.Notes Phys.} {\bf 871} (2013) 51--86, [\href{http://arxiv.org/abs/1208.0536}{{\tt arXiv:1208.0536}}].

\bibitem{Bigazzi:2014qsa}
F.~Bigazzi and A.~L. Cotrone, {\it {Holographic QCD with Dynamical Flavors}},  {\em JHEP} {\bf 01} (2015) 104, [\href{http://arxiv.org/abs/1410.2443}{{\tt arXiv:1410.2443}}].

\bibitem{Li:2015uea}
S.-w. Li, A.~Schmitt, and Q.~Wang, {\it {From holography towards real-world nuclear matter}},  {\em Phys. Rev.} {\bf D92} (2015), no.~2 026006, [\href{http://arxiv.org/abs/1505.04886}{{\tt arXiv:1505.04886}}].

\bibitem{BitaghsirFadafan:2018uzs}
K.~Bitaghsir~Fadafan, F.~Kazemian, and A.~Schmitt, {\it {Towards a holographic quark-hadron continuity}},  {\em JHEP} {\bf 03} (2019) 183, [\href{http://arxiv.org/abs/1811.08698}{{\tt arXiv:1811.08698}}].

\bibitem{Kovensky:2021kzl}
N.~Kovensky, A.~Poole, and A.~Schmitt, {\it {Building a realistic neutron star from holography}},  {\em Phys. Rev. D} {\bf 105} (2022), no.~3 034022, [\href{http://arxiv.org/abs/2111.03374}{{\tt arXiv:2111.03374}}].

\bibitem{Bartolini:2022gdf}
L.~Bartolini and S.~B. Gudnason, {\it {Symmetry energy in holographic QCD}},  {\em SciPost Phys.} {\bf 16} (2024) 156, [\href{http://arxiv.org/abs/2209.14309}{{\tt arXiv:2209.14309}}].

\bibitem{Bartolini:2023wis}
L.~Bartolini and S.~B. Gudnason, {\it {Neutron stars in the Witten-Sakai-Sugimoto model}},  {\em JHEP} {\bf 11} (2023) 209, [\href{http://arxiv.org/abs/2307.11886}{{\tt arXiv:2307.11886}}].

\bibitem{Kovensky:2019bih}
N.~Kovensky and A.~Schmitt, {\it {Heavy Holographic QCD}},  {\em JHEP} {\bf 02} (2020) 096, [\href{http://arxiv.org/abs/1911.08433}{{\tt arXiv:1911.08433}}].

\bibitem{Kovensky:2020xif}
N.~Kovensky and A.~Schmitt, {\it {Holographic quarkyonic matter}},  {\em JHEP} {\bf 09} (2020) 112, [\href{http://arxiv.org/abs/2006.13739}{{\tt arXiv:2006.13739}}].

\bibitem{Kovensky:2021ddl}
N.~Kovensky and A.~Schmitt, {\it {Isospin asymmetry in holographic baryonic matter}},  {\em SciPost Phys.} {\bf 11} (2021), no.~2 029, [\href{http://arxiv.org/abs/2105.03218}{{\tt arXiv:2105.03218}}].

\bibitem{Kovensky:2023mye}
N.~Kovensky, A.~Poole, and A.~Schmitt, {\it {Phases of cold holographic QCD: Baryons, pions and rho mesons}},  {\em SciPost Phys.} {\bf 15} (2023), no.~4 162, [\href{http://arxiv.org/abs/2302.10675}{{\tt arXiv:2302.10675}}].

\bibitem{Aharony:2008an}
O.~Aharony and D.~Kutasov, {\it {Holographic Duals of Long Open Strings}},  {\em Phys. Rev.} {\bf D78} (2008) 026005, [\href{http://arxiv.org/abs/0803.3547}{{\tt arXiv:0803.3547}}].

\bibitem{Hashimoto:2008sr}
K.~Hashimoto, T.~Hirayama, F.-L. Lin, and H.-U. Yee, {\it {Quark Mass Deformation of Holographic Massless QCD}},  {\em JHEP} {\bf 07} (2008) 089, [\href{http://arxiv.org/abs/0803.4192}{{\tt arXiv:0803.4192}}].

\bibitem{Argyres:2008sw}
P.~C. Argyres, M.~Edalati, R.~G. Leigh, and J.~F. Vazquez-Poritz, {\it {Open Wilson Lines and Chiral Condensates in Thermal Holographic QCD}},  {\em Phys. Rev.} {\bf D79} (2009) 045022, [\href{http://arxiv.org/abs/0811.4617}{{\tt arXiv:0811.4617}}].

\bibitem{Casero:2007ae}
R.~Casero, E.~Kiritsis, and A.~Paredes, {\it {Chiral symmetry breaking as open string tachyon condensation}},  {\em Nucl. Phys. B} {\bf 787} (2007) 98--134, [\href{http://arxiv.org/abs/hep-th/0702155}{{\tt hep-th/0702155}}].

\bibitem{Bergman:2007pm}
O.~Bergman, S.~Seki, and J.~Sonnenschein, {\it {Quark mass and condensate in HQCD}},  {\em JHEP} {\bf 12} (2007) 037, [\href{http://arxiv.org/abs/0708.2839}{{\tt arXiv:0708.2839}}].

\bibitem{Dhar:2007bz}
A.~Dhar and P.~Nag, {\it {Sakai-Sugimoto model, Tachyon Condensation and Chiral symmetry Breaking}},  {\em JHEP} {\bf 01} (2008) 055, [\href{http://arxiv.org/abs/0708.3233}{{\tt arXiv:0708.3233}}].

\bibitem{Dhar:2008um}
A.~Dhar and P.~Nag, {\it {Tachyon condensation and quark mass in modified Sakai-Sugimoto model}},  {\em Phys. Rev. D} {\bf 78} (2008) 066021, [\href{http://arxiv.org/abs/0804.4807}{{\tt arXiv:0804.4807}}].

\bibitem{McNees:2008km}
R.~McNees, R.~C. Myers, and A.~Sinha, {\it {On quark masses in holographic QCD}},  {\em JHEP} {\bf 11} (2008) 056, [\href{http://arxiv.org/abs/0807.5127}{{\tt arXiv:0807.5127}}].

\bibitem{Jarvinen:2011qe}
M.~J{\"a}rvinen and E.~Kiritsis, {\it {Holographic Models for QCD in the Veneziano Limit}},  {\em JHEP} {\bf 03} (2012) 002, [\href{http://arxiv.org/abs/1112.1261}{{\tt arXiv:1112.1261}}].

\bibitem{Alho:2012mh}
T.~Alho, M.~J\"arvinen, K.~Kajantie, E.~Kiritsis, and K.~Tuominen, {\it {On finite-temperature holographic QCD in the Veneziano limit}},  {\em JHEP} {\bf 01} (2013) 093, [\href{http://arxiv.org/abs/1210.4516}{{\tt arXiv:1210.4516}}].

\bibitem{Albrecht:2010eg}
D.~Albrecht and J.~Erlich, {\it {Pion condensation in holographic QCD}},  {\em Phys. Rev. D} {\bf 82} (2010) 095002, [\href{http://arxiv.org/abs/1007.3431}{{\tt arXiv:1007.3431}}].

\bibitem{Nishihara:2014nva}
H.~Nishihara and M.~Harada, {\it {Enhancement of Chiral Symmetry Breaking from the Pion condensation at finite isospin chemical potential in a holographic QCD model}},  {\em Phys. Rev. D} {\bf 89} (2014), no.~7 076001, [\href{http://arxiv.org/abs/1401.2928}{{\tt arXiv:1401.2928}}].

\bibitem{Cao:2020ske}
X.~Cao, H.~Liu, D.~Li, and G.~Ou, {\it {QCD phase diagram at finite isospin chemical potential and temperature in an IR-improved soft-wall AdS/QCD model}},  {\em Chin. Phys. C} {\bf 44} (2020), no.~8 083106, [\href{http://arxiv.org/abs/2001.02888}{{\tt arXiv:2001.02888}}].

\bibitem{Adhikari:2020kdn}
P.~Adhikari, J.~O. Andersen, and M.~A. Mojahed, {\it {Condensates and pressure of two-flavor chiral perturbation theory at nonzero isospin and temperature}},  {\em Eur. Phys. J. C} {\bf 81} (2021), no.~2 173, [\href{http://arxiv.org/abs/2010.13655}{{\tt arXiv:2010.13655}}].

\bibitem{Gell-Mann:1968hlm}
M.~Gell-Mann, R.~J. Oakes, and B.~Renner, {\it {Behavior of current divergences under SU(3) x SU(3)}},  {\em Phys. Rev.} {\bf 175} (1968) 2195--2199.

\bibitem{Callebaut:2011ab}
N.~Callebaut, D.~Dudal, and H.~Verschelde, {\it {Holographic rho mesons in an external magnetic field}},  {\em JHEP} {\bf 03} (2013) 033, [\href{http://arxiv.org/abs/1105.2217}{{\tt arXiv:1105.2217}}].

\bibitem{Aharony:2006da}
O.~Aharony, J.~Sonnenschein, and S.~Yankielowicz, {\it {A Holographic model of deconfinement and chiral symmetry restoration}},  {\em Annals Phys.} {\bf 322} (2007) 1420--1443, [\href{http://arxiv.org/abs/hep-th/0604161}{{\tt hep-th/0604161}}].

\bibitem{Horigome:2006xu}
N.~Horigome and Y.~Tanii, {\it {Holographic chiral phase transition with chemical potential}},  {\em JHEP} {\bf 01} (2007) 072, [\href{http://arxiv.org/abs/hep-th/0608198}{{\tt hep-th/0608198}}].

\bibitem{progress}
N.~Kovensky and A.~Schmitt. Work in Progress.

\bibitem{Kovensky:2021wzu}
N.~Kovensky, A.~Poole, and A.~Schmitt, {\it {Predictions for neutron stars from holographic nuclear matter}},  {\em SciPost Phys. Proc.} {\bf 6} (2022) 019, [\href{http://arxiv.org/abs/2112.10633}{{\tt arXiv:2112.10633}}].

\bibitem{Colangelo:2003hf}
G.~Colangelo and S.~D{\"u}rr, {\it {The Pion mass in finite volume}},  {\em Eur. Phys. J. C} {\bf 33} (2004) 543--553, [\href{http://arxiv.org/abs/hep-lat/0311023}{{\tt hep-lat/0311023}}].

\bibitem{Adhikari:2020ufo}
P.~Adhikari and J.~O. Andersen, {\it {Quark and pion condensates at finite isospin density in chiral perturbation theory}},  {\em Eur. Phys. J. C} {\bf 80} (2020), no.~11 1028, [\href{http://arxiv.org/abs/2003.12567}{{\tt arXiv:2003.12567}}].

\bibitem{Chiba:2023ftg}
R.~Chiba and T.~Kojo, {\it {Sound velocity peak and conformality in isospin QCD}},  {\em Phys. Rev. D} {\bf 109} (2024), no.~7 076006, [\href{http://arxiv.org/abs/2304.13920}{{\tt arXiv:2304.13920}}].

\bibitem{Graf:2015tda}
T.~Graf, J.~Schaffner-Bielich, and E.~S. Fraga, {\it {The impact of quark masses on pQCD thermodynamics}},  {\em Eur. Phys. J. A} {\bf 52} (2016), no.~7 208, [\href{http://arxiv.org/abs/1507.08941}{{\tt arXiv:1507.08941}}].

\bibitem{Tews:2018kmu}
I.~Tews, J.~Carlson, S.~Gandolfi, and S.~Reddy, {\it {Constraining the speed of sound inside neutron stars with chiral effective field theory interactions and observations}},  {\em Astrophys. J.} {\bf 860} (2018), no.~2 149, [\href{http://arxiv.org/abs/1801.01923}{{\tt arXiv:1801.01923}}].

\bibitem{Evans:2011eu}
N.~Evans, A.~Gebauer, M.~Magou, and K.-Y. Kim, {\it {Towards a Holographic Model of the QCD Phase Diagram}},  {\em J. Phys.} {\bf G39} (2012) 054005, [\href{http://arxiv.org/abs/1109.2633}{{\tt arXiv:1109.2633}}].

\bibitem{Rougemont:2023gfz}
R.~Rougemont, J.~Grefa, M.~Hippert, J.~Noronha, J.~Noronha-Hostler, I.~Portillo, and C.~Ratti, {\it {Hot QCD phase diagram from holographic Einstein\textendash{}Maxwell\textendash{}Dilaton models}},  {\em Prog. Part. Nucl. Phys.} {\bf 135} (2024) 104093, [\href{http://arxiv.org/abs/2307.03885}{{\tt arXiv:2307.03885}}].

\bibitem{Adhikari:2018cea}
P.~Adhikari, J.~O. Andersen, and P.~Kneschke, {\it {Pion condensation and phase diagram in the Polyakov-loop quark-meson model}},  {\em Phys. Rev. D} {\bf 98} (2018), no.~7 074016, [\href{http://arxiv.org/abs/1805.08599}{{\tt arXiv:1805.08599}}].

\bibitem{Chiba:2024cny}
R.~Chiba, T.~Kojo, and D.~Suenaga, {\it {Thermal effects on sound velocity peak and conformality in isospin QCD}},  \href{http://arxiv.org/abs/2403.02538}{{\tt arXiv:2403.02538}}.

\end{thebibliography}\endgroup

\end{document}